\documentclass[%
reprint,
superscriptaddress,
showpacs,
 amsmath,amssymb,
 aps,
 pre,
longbibliography
]{revtex4-1}

\usepackage{psfrag,graphicx,epsfig}
\usepackage[usenames,dvipsnames,svgnames,tables]{xcolor}
\usepackage{subfigure}
\usepackage{graphicx}

\renewcommand{\vec}[1]{\mathbf{#1}}

\newcommand{\tr}{\textrm{tr}}

\newcommand{\dissip}{\varepsilon}

\begin{document}

\title{Irreversibility and small-scale generation in 3D turbulent flows}

\author{Alain Pumir}
\email{alain.pumir@ens-lyon.fr, hxu@tsinghua.edu.cn}
\affiliation{Ecole Normale Sup\'erieure de Lyon, 69007 Lyon, France}
\affiliation{Max Planck Institute for Dynamics and Self-Organization (MPIDS), 37077 G\"ottingen, Germany}

\author{Haitao Xu}
\email{alain.pumir@ens-lyon.fr, hxu@tsinghua.edu.cn}
\affiliation{Max Planck Institute for Dynamics and Self-Organization (MPIDS), 37077 G\"ottingen, Germany}
\affiliation{Center for Combustion Energy and Department of Thermal Engineering, Tsinghua University, 100084, Beijing, China}

\author{Rainer Grauer}
\affiliation{Institute for Theoretical Physics I, Ruhr Universit\"at Bochum, 44780 Bochum, Germany}

\author{Eberhard Bodenschatz}
\affiliation{Max Planck Institute for Dynamics and Self-Organization (MPIDS), 37077 G\"ottingen, Germany}
\affiliation{Institute for Nonlinear Dynamics, University of G\"ottingen, 37077 G\"ottingen, Germany}
\affiliation{Laboratory of Atomic and Solid State Physics and Sibley School of Mechanical and Aerospace Engineering, Cornell University, Ithaca, NY 14853, USA}

%**************************************************************************

\begin{abstract}
In three-dimensional turbulent flows energy is supplied at 
large scales and cascades down to the smallest scales where
viscosity dominates.
The flux of energy through scales implies the generation of small scales
from larger ones, which is the fundamental reason for the irreversibility 
of the dynamics of turbulent flows.
As we showed recently, this irreversibility manifests itself by an asymmetry of the probability distribution
of the instantaneous power $p$
of the forces acting on fluid elements. 
In particular, the third moment of $p$ was found to be negative.
Yet, a physical connection between the irreversibility manifested in the distribution of 
$p$ and the energy flux or small-scale generation in turbulence has not been established.
Here, with analytical calculations and support from numerical 
simulations of fully developed turbulence, we connect the asymmetry in the 
power distribution, {\it i.e.},  
the negative value of  $\langle p^3 \rangle$, to
the generation of small scales, or more precisely, to the 
amplification (stretching) of vorticity in turbulent flows.
Our result is the first step towards a quantitative understanding of the origin of the irreversibility
observed at the level of individual Lagrangian trajectories in turbulent flows.
\end{abstract}

\maketitle

\section{Introduction}
\label{sec:intro}

The generation of small scales, or large velocity gradients, 
is one of the most striking
physical phenomenon of 3-dimensional (3D) turbulent fluid flows,
and is responsible for a flux of energy $\dissip$ from large to small scales.
Remarkably, in the limit of very small viscosity or
very large Reynolds number, 
the third moment of the longitudinal velocity difference
between two points separated by a distance $x$, 
$\langle \Delta u(x)^3 \rangle$, is related to the energy flux
by the relation
$\langle \Delta u(x)^3 \rangle = - \frac{4}{5} \dissip x $,
which is one of the very few exact results in turbulence theory~\cite{K41}.
In elementary terms, two points are more likely to be pushed closer
together (repelled) 
when their relative energy is large (small)~\cite{FGV:2001}. 
This fundamental
asymmetry persists all the way down to very small distances so the 
third moment of the velocity derivative $\partial_x u_x$
is {\it negative}: $\langle ( \partial_x u_x)^3 \rangle \le 0$. 
In fact, available
data from experiments using hot-wire 
anemometry~\cite{BT47,Townsend51} and from direct numerical simulations 
(DNS) have led to the conclusion that the normalized third moment of
$\partial_x u_x$, i.e., the skewness, 
$S_{\partial_x u_x} \equiv \langle ( \partial_x u_x )^3 \rangle / \langle (\partial_x u_x)^2 \rangle^{3/2}  $,
is negative  and approximately $ -0.5$, with at most a weak dependence 
on the Reynolds number~\cite{SA97,Ishihara07}.
In homogeneous isotropic flows, the seminal work of 
Betchov~\cite{Betchov56} shows that the third moment of $\partial_x u_x$ 
is related to the generation of small scales in turbulence, 
through amplification of vorticity by vortex stretching.

Because of the existence of an energy flux from large to small scales,
turbulence is a non-equilibrium phenomenon, thus intrinsically
irreversible.  
The possibility to probe turbulence by following the motion of
individual particles in both numerical and laboratory high-Reynolds-number 
flows~\cite{YP89,LVC+01,MMMP01},
leads to new insights on irreversibility and offers new
opportunities for quantitatively understanding turbulence~\cite{BOX+06,SY11,FXP+12}.

Recently, we observed that the energy differences along particle 
trajectories present an intriguing asymmetry: kinetic energy grows 
more slowly than it drops along a trajectory~\cite{XPFB14}. 
The consequence of this asymmetry
is that the third moment 
of the power $p = \mathbf{a} \cdot \mathbf{u}$ is {\it negative},
where 
$\vec{u} $ and $\vec{a} $ are the velocity and acceleration of the fluid
(see \cite{Leveque14} for a related discussion).
As a possible explanation, one may expect the pressure gradient, which dominates
the fluctuations of the power, to provide an explanation for the
negative sign of the third moment of $p$~\cite{PXB+14,OY51,VY:1999}.
Unexpectedly, however, in 3D, the contribution of the pressure gradient 
to the third moment of power is very small~\cite{PXB+14}.

Here, we
provide a physical relation between 
the negative third moment of $p$ and the generation of small scales 
by turbulence, i.e., vortex stretching.
In the following, it is convenient to decompose the power as
\begin{equation}
p = p_L + p_C
\label{eq:p_decomp}
\end{equation}
where $p_L = \mathbf{u} \cdot \mathbf{a}_L = \mathbf{u} \cdot \partial_t \mathbf{u}$
and $ p_C = \mathbf{u} \cdot \mathbf{a}_C  = \mathbf{u} \cdot  ( \mathbf{u} \cdot \nabla )  \mathbf{u} $
are the local and convective parts, respectively.
We find that 
the magnitude of $p = p_L + p_C$ is much smaller than the magnitudes of 
its components 
$p_L$ and $p_C $, which implies significant cancellation between 
$p_C$ and $p_L$. 
On average, the magnitude of $p_C$ is larger than that of 
$p_L$.
Note however that the
cancellation between $p_L$ and $p_C$ does not automatically
follow from the well-known cancellation between $\vec{a}_L$ and $\vec{a}_C$
~\cite{Tennekes71,Tsin:2001,Gulitski:2007b}, since $p_{L}$, $p_C$ involve
only one component of $\mathbf{a}_{L}$, $\mathbf{a}_C$.
We demonstrate that the moments of $p$, up to the third order, are dominated by the moments of $p_C$. 
In particular, the third moment $\langle p^3 \rangle$ has the same sign as $\langle p_C^3 \rangle$.
We show analytically that 
$\langle p_C^3 \rangle$ is a surrogate for
vorticity amplification.
This, together with the observation that $\langle p_C^3 \rangle $ determines the sign of $\langle p^3 \rangle$,
leads us to the conclusion 
that the origin of the negative sign of the third moment 
$\langle p^3 \rangle$
comes in fact from small scale generation, 
thus clearly establishing a relation between the generation of small scales
and the observed irreversibility in the flow.

\section{Numerical methods}
\label{sec:num}

\subsection{Direct Numerical Simulation of Navier-Stokes Turbulence}
\label{sec:dns_a}

We investigated numerically turbulent flows, obtained by solving directly
the Navier--Stokes equations:
\begin{eqnarray}
\partial_t \vec{u}( \vec{x}, t) & + & ( \vec{u}(\vec{x} ,t ) \cdot \nabla ) \vec{u} (\vec{x} ,t ) \nonumber \\
& = &- \nabla P (\vec{x}, t) + \nu \nabla^2 \vec{u}(\vec{x}, t) + \vec{f} ( \vec{x} , t) \label{eq:NS} \\
\nabla \cdot \vec{u} (\vec{x} , t) & = & 0 \label{eq:incomp}
\end{eqnarray}
where $\vec{u}(\vec{x} ,t )$ denotes the Eulerian velocity field, $P$ is the pressure, $\nu$ is the viscosity,
and $\vec{f}(\vec{x} ,t )$ is a forcing term; the mass density is arbitrarily set to unity.
Solving the equations in a simple cubic box of size $(2\pi)^3$ with periodic
boundary conditions allows us to use efficient pseudo-spectral methods.

The forcing term acts at large scales, or equivalently, on Fourier modes at 
low wavenumbers, $| \vec{k} | \le K_f$.
It is adjusted according to a method proposed in~\cite{lamorgese05}, in such 
a way that the injection rate of energy, $\dissip_i$, remains constant:
\begin{equation}
\vec{f}_{\vec{k}} = \dissip_i ~ \frac{ \vec{u}_{\vec{k}} }{\sum_{|\vec{k}| \le K_f}| \vec{u}_{\vec{k}} |^2 }
\quad \mathrm{if} \quad | \vec{k} | \le K_f
\ ,
\label{eq:forcing}
\end{equation}
with $K_f = 1.5$.
In the code units, the energy injection rate $\dissip_i$ has been set to 
$\dissip_i = 10^{-3}$. 
Note that in stationary turbulent flows, the energy injection rate equals the energy dissipation rate, $\dissip_i = \dissip$.

The code is fully dealiased, using the $2/3$-rule method~\cite{Orszag:71}.
We have chosen two different resolutions, corresponding to
the highest resolved wavenumber of $k_{\max} = 256$ and $k_{\max} = 384$ 
(effectively equivalent to $768$ and
$1152$ grid points in each spatial direction), with the corresponding 
values of the viscosity $\nu = 1.6 \times 10^{-4}$ and 
$ 9.0 \times 10^{-5}$, respectively.
With these values, the Kolmogorov scale $\eta = (\nu^3/\dissip)^{1/4}$
is such that the product $k_{\max} \times \eta $ is very close to $2$ in both cases, 
ensuring adequate spatial resolution.
The corresponding Reynolds numbers are $R_\lambda \approx 193$ and $275$, respectively.

Once expressed in terms of spatial modes, Eq.~(\ref{eq:NS}) reduces to a 
large set of ordinary differential
equations, which were integrated
using the second-order Adams-Bashforth scheme.
The time step $\delta t$ has been chosen so that the Courant number
$\mathrm{Co}= u_\mathrm{rms} \cdot k_\mathrm{max} \delta t \lesssim 0.1$, where
$u_\mathrm{rms}$ is the root mean square value of one component of velocity.

\subsection{Data from the Johns Hopkins University Database}
\label{sec:dns_jhu}

We also used additional numerical simulation data at $R_\lambda = 430$ from the Turbulence Database 
of the Johns Hopkins University.
The flow is documented in~\cite{li2008}.
We computed the statistics presented here with at the minimum $2 \times 10^8$ 
points. 

\begin{table}[bt]
\begin{center}
\begin{tabular}{|c||c|c|c|}
\hline
$R_{\lambda}$ & $193$ & $275$ & $430$ \\ 
\hline
$\langle p^2 \rangle/\dissip^2$ & $3.83 \times 10^{2}$ & $7.36  \times 10^{2}$ & $1.32 \times 10^3$ \\
$\langle p_C^2 \rangle/\dissip^2$ & $2.15 \times 10^{3}$ & $5.00  \times 10^{3}$ & $1.20 \times 10^4$ \\
$\langle p_L^2 \rangle/\dissip^2$ & $1.78 \times 10^{3}$ & $4.25  \times 10^{3}$ & $1.07 \times 10^4$ \\
$-\langle p_L ~  p_C \rangle/\dissip^2$ & $1.77 \times 10^{3}$ & $4.26  \times 10^{3}$ & $1.07  \times 10^4$ \\
\hline
$15 \langle p_C^2 \rangle/(\dissip^2 R_\lambda^2)$ & $0.87$ & $0.99$ & $0.96$ \\
\hline
$\beta$ & $0.83$ & $0.86$ & $0.90$ \\  
\hline
\end{tabular}
\end{center}
\caption{
Second moments of the distributions of 
$p/\dissip$, $p_C/\dissip$ and $p_L/\dissip$
at the three Reynolds numbers studied in this article. 
The correlation coefficient between $p_C$ and $p_L$ is 
approaching $-1$ as Reynolds number increases.
The values of $\beta$ are measured from fitting the conditional averages $\langle p_L | p_C \rangle = - \beta p_C$.
}
\label{table:mom2}
\end{table}

\begin{table}[bt]
\begin{center}
\begin{tabular}{|c||c|c|c|}
\hline
$R_{\lambda}$ & $193$ & $275$ & $430$ \\ 
\hline
$-\langle p^3 \rangle/\dissip^3$ & $3.87 \times 10^{3}$ & $1.23  \times 10^{4}$ & $3.21 \times 10^{4}$ \\
$-\langle p_C^3 \rangle/\dissip^3$ & $5.39 \times 10^{4}$ & $2.40  \times 10^{5}$ & $1.00 \times 10^6$ \\
$\langle p_C^2 \,  p_L \rangle/\dissip^3 $ & $4.54 \times 10^{4}$ & $2.05  \times 10^{5}$ & $8.99 \times 10^5$ \\
$-\langle p_C \,  p_L^2 \rangle/\dissip^3$ & $4.02 \times 10^{4}$ & $1.84  \times 10^{5}$ & $8.29 \times 10^5$ \\
$\langle p_L^3 \rangle/\dissip^3 $ & $3.44 \times 10^{4}$ & $1.63  \times 10^{5}$ & $7.63 \times 10^{5}$ \\
\hline
$\zeta = \langle p_L^2 p \rangle / \langle p_C^3 \rangle$ & $0.108$ & $0.088$ & $0.066$ \\
$1-\beta$ & $0.17$ & $0.14$ & $0.11$ \\
\hline
$\langle p^3 \rangle / \langle p_C^3 \rangle$ & $0.072$ & $0.051$ & $0.032$ \\  
$1- \beta - \zeta $ & $0.061$ & $0.052$ & $0.037$ \\
\hline
$- \langle p_L^3 \rangle / \langle p_C^3 \rangle$ & $0.64$ & $0.68$ & $0.76$ \\  
$\beta - 2\zeta $ & $0.62$ & $0.69$ & $0.77$ \\
\hline
$\langle p_C \, p_L^2 \rangle / \langle p_C^3 \rangle$ & $0.75$ & $0.77$ & $0.83$ \\  
$\beta - \zeta $ & $0.72$ & $0.77$ & $0.83$ \\
\hline
$-\langle p_C^2 \, p_L \rangle / \langle p_C^3 \rangle$ & $0.84$ & $0.85$ & $0.90$ \\  
$\beta$ & $0.83$ & $0.86$ & $0.90$ \\
\hline
\end{tabular}
\caption{
Third moments of the distributions of $p/\dissip$, $p_C/\dissip$ and $p_L/\dissip$
at the three Reynolds numbers studied in this article. 
The last 8 rows compare the normalized moments with our predictions.
}
\label{table:mom3}
\end{center}
\end{table}

\section{Theoretical background}

\subsection{Elementary relations}
To investigate the moments of $p$, $p_C$ and $p_L$, we first note that 
$p_C$ reduces to a 
simple form that is particularly useful, namely:
\begin{equation}
p_C = \mathbf{u} \cdot ( \mathbf{u} \cdot \nabla ) \mathbf{u} = \mathbf{u} 
\cdot \mathbf{S} \cdot \mathbf{u} \, ,
\label{eq:p_C_uSu}
\end{equation}
where the rate of strain tensor $\mathbf{S} $ is the symmetric part of
the velocity gradient tensor $\nabla \mathbf{u}$:
$\mathbf{S} = [\nabla \mathbf{u} + (\nabla \mathbf{u})^{T}]/2$ or 
$S_{ij} = (\partial_i u_j + \partial_j u_i)/2$.
Geometrically, the straining motion decomposes into
a superposition of compression or stretching along three orthogonal
directions, denoted by $\mathbf{e}_i$, with three straining rates, 
$\lambda_i$. 
The vectors $\mathbf{e}_i$ and the straining rates $\lambda_i$ are
the eigenvectors and eigenvalues of $\mathbf{S}$. 
A positive (respectively negative) value of $\lambda_i$
corresponds to stretching (respectively compression) in the direction 
$\mathbf{e}_i$. 
Volume conservation (incompressibility) imposes that 
$\lambda_1 + \lambda_2 + \lambda_3 = 0$, i.e.,
the amount of stretching and compression along the
three directions $\vec{e}_i$ sums up to $0$.

Equation \eqref{eq:p_C_uSu} shows that in a steady (frozen) flow, 
the kinetic energy of a fluid element changes only through the
action of the rate of strain.
The antisymmetric part of the velocity gradient tensor
expresses the local rotation rate in the fluid, and is characterized 
by the vorticity $\mathbf{\omega}$.
The expression for the amplification (stretching) of vorticity in the flow
is given by 
$\langle \omega \cdot \mathbf{S} \cdot \omega \rangle$~\cite{Tennekes,frisch95}.
In a statistically homogeneous flow, the following identity holds:
$\langle \omega \cdot \mathbf{S} \cdot \omega \rangle = - \frac{4}{3} 
\langle \tr(\mathbf{S}^3) \rangle$~\cite{Betchov56}. 
Last, we note that in a homogeneous isotropic flow, the second and third moments
of $\partial_x u_x$ can be simply expressed in terms of the moments of 
$\tr(\mathbf{S}^{2})$:
$\langle (\partial_x u_x)^2 \rangle = \frac{2}{15} \langle \tr(\mathbf{S}^2) \rangle$,
and $\tr(\mathbf{S}^3 )$:
$\langle (\partial_x u_x)^3 \rangle = \frac{8}{35} \langle \tr(\mathbf{S}^3) \rangle$~\cite{Betchov56}.

\subsection{Decomposition of power $p$: order of magnitudes}
\label{sec:decomp_p}

The magnitudes of the fluctuations of the convective and local components
of $p$ may be estimated from simple dimensional arguments:
$|p_C| \sim |p_L| \sim U^2/\tau_K$, where $U$ is the typical size of the 
velocity fluctuations, and $\tau_K$ is the fastest time scales of the 
turbulent eddies. Using the known relation 
$\tau_K \sim (U^2 / \dissip) / R_\lambda$~\cite{Tennekes,frisch95}, one finds
$|p_C| \sim | p_L | \sim \dissip R_\lambda$, where $R_\lambda$ is the
Reynolds number based on the Taylor microscale, and characterizes the intensity
of turbulence.
The growth of the variances of $p_C/\dissip$ and $p_L/\dissip$
as $R_\lambda^2$, as predicted by this simple dimensional argument, 
is found to be consistent with our DNS results, see
Table~\ref{table:mom2}.
This result sharply contrasts with the fact that the variance of $p$
is known to grow more slowly
with the Reynolds number, as $R_\lambda^{4/3}$~\cite{XPFB14}.
This difference in the observed scalings as a function of the Reynolds 
number is due to a very strong cancellation between $p_L$ and $p_C$,
see Table~\ref{table:mom2}. 
We observe that the magnitudes of the third moments 
$\langle p_C^m p_L^n \rangle$, with $m + n = 3$, are found to increase with 
$m$, see Table~\ref{table:mom3}, signaling that the contribution of
$p_C$ to the third moments is more significant than that of $p_L$.
In fact, as we will show,  
the sign of $\langle p^3 \rangle$ is dominated by $\langle p_C^3 \rangle$.

Although the cancellation between $p_C$ and $p_L$ is reminiscent of 
the well-documented
cancellation between $\mathbf{a}_C$ and 
$\mathbf{a}_L$~\cite{Tennekes71,Tsin:2001,Gulitski:2007b}, we stress
that it {\it cannot} be 
deduced from the results 
of~\cite{Tsin:2001,Gulitski:2007b}.
In fact, $p_C$ and $p_L$ involve the
projections along the direction of the velocity $\mathbf{u}$,
of $\mathbf{a}_C$ and $\mathbf{a}_L$, respectively. 
Our results therefore show that the cancellation
between $\mathbf{a}_C$ and $\mathbf{a}_L$ affects their components along 
the velocity direction, which does not result automatically from
~\cite{Tsin:2001,Gulitski:2007b}.

In the following subsections, we begin by expressing 
$\langle p_C^3 \rangle$ 
in terms of vortex stretching, before establishing 
the prevalence of $p_C$ on $\langle p^3 \rangle$.

\section{Results }

\subsection{Vortex stretching and moments of $p_C$}
\label{sec:3rd_mom_pC}

It is convenient to express $p_C$ (Eq.~\eqref{eq:p_C_uSu}) by projecting
the velocity $\mathbf{u}$ and the rate of strain $\mathbf{S}$ in the basis
of the three perpendicular unit vectors $\mathbf{e}_i$ characterizing the 
straining motion.
In this basis, the velocity $\mathbf{u}$ is decomposed as:
$\mathbf{u} = \sum_{i=1}^{3} u_i \mathbf{e}_i$, 
where $u_i = \vec{u} \cdot \vec{e}_i$ is the coordinate of 
$\vec{u}$ along the direction $\vec{e}_i$, and the rate of strain tensor 
is expressed as 
$\mathbf{S} = \sum_{i=1}^{3} \lambda_i \mathbf{e}_i \mathbf{e}_i$.
Denoting $\hat{x}_i$ the cosines of the angles between the velocity 
$\mathbf{u}$ and the unit vectors $\mathbf{e}_i$: 
$\hat{x}_i \equiv \vec{u} \cdot \vec{e}_i / | \vec{u} | = u_i / | \vec{u} |$, 
the expression of $p_C$ reduces to: 
\begin{equation}
p_C = \sum_{i=1}^3 \lambda_i u_i^2  = \mathbf{u}^2 ~ \sum_{i=1}^3 \lambda_i \hat{x}_i^2 .
\label{eq:pC_eigenb}
\end{equation}

In a turbulent velocity field, small wave numbers (or large scales)
provide the main 
contribution to the velocity field, $\mathbf{u}$, whereas the rate of strain
$\mathbf{S}$ is determined by the large wave numbers (or small scales). 
The two fields $\vec{u}$ and $\vec{S}$ are therefore expected to be
only weakly correlated. 
Let us now assume that $\mathbf{S}$ and $\mathbf{u}$ 
are  uncorrelated. This approximation implies that the three
cosines, $\hat{x}_i$, are uniformly distributed between $-1$ and $1$.
Geometrically, the three cosines are the coordinates of 
a point that is uniformly distributed on the unit sphere in 3D. 
This assumption allows us to compute the averages necessary to 
evaluate explicitly 
the third moment of $p_C$.

Namely, Eq.~\eqref{eq:pC_eigenb} leads to: 
\begin{eqnarray}
\langle p_C^3 \rangle & = & 
\left\langle \left( |\vec{u}|^2 \sum_{i=1}^3 \lambda_i \hat{x}_i^2 \right)^3 \right\rangle \nonumber \\
& = & \left\langle |\mathbf{u}|^6 \right\rangle \left\langle \left(\sum_{i=1}^3 \lambda_i \hat{x}_i^2 \right)^3 \right\rangle .
\label{eq:decomp_p_C}
\end{eqnarray}
The assumption that $\vec{u}$ and $\mathbf{S}$ are uncorrelated 
also implies that all the cosines $\hat{x}_i$, ($i=1 \ldots 3$), 
are independent of
the eigenvalues of $\mathbf{S}$. 
As a consequence,
$\langle \lambda_i^m \hat{x}_i^n \rangle = \langle \lambda_i^m \rangle \langle \hat{x}_i^n \rangle$ for any $m$ and $n$. 
Using the observation that
$( \hat{x}_1 , \hat{x}_2 , \hat{x}_3 )$ represents the coordinates of 
a point that is uniformly distributed on the unit sphere, which 
gives the symmetry relations such as $\langle \hat{x}_1^6 \rangle = \langle \hat{x}_2^6 \rangle = \langle \hat{x}_3^6 \rangle$,
$\langle \hat{x}_1^4 \hat{x}_2^2 \rangle = \langle \hat{x}_2^4 \hat{x}_3^2 \rangle = \langle \hat{x}_3^4 \hat{x}_1^2 \rangle$, etc, 
we therefore obtain
\begin{eqnarray}
\left\langle \left(\sum_{i=1}^3 \lambda_i \hat{x}_i^2 \right)^3 \right\rangle
& = &  \left\langle \hat{x}_1^6 \right\rangle \left\langle \sum_{i=1}^3 \lambda_i^3 \right\rangle \nonumber \\
& & + 3 \left\langle \hat{x}_1^4 \hat{x}_2^2 \right\rangle \left\langle \sum_{i,j = 1, i\ne j}^3 \lambda_i^2 \lambda_j \right\rangle  \nonumber \\
& &   + 6 \left\langle \hat{x}_1^2 \hat{x}_2^2 \hat{x}_3^2 \right\rangle \left\langle \lambda_1 \lambda_2 \lambda_3 \right\rangle . 
\label{eq:pC_thrd_dec_1}
\end{eqnarray}
The averages of the products of $\hat{x}_i$ in Eq.~\eqref{eq:pC_thrd_dec_1} can
be calculated by using elementary geometrical 
considerations~\cite{Betchov56} and the results are:
\begin{equation}
\langle {\hat x}_1^6\rangle = \frac{1}{7}~,~~~
\langle {\hat x}_1^4 {\hat x}_2^2 \rangle = \frac{1}{35}~,~~~
\langle {\hat x}_1^2 {\hat x}_2^2 {\hat x}_3^2 \rangle = \frac{1}{105} .
\label{eq:sphere_avg}
\end{equation} 
Substituting Eqs.~\eqref{eq:sphere_avg} and \eqref{eq:pC_thrd_dec_1} into 
Eq.~\eqref{eq:decomp_p_C}, 
and using the incompressibility of the flow, $\lambda_1 + \lambda_2 + \lambda_3 =0$, 
leads to the following expression for the third moment of $p_C$:
\begin{eqnarray}
\langle p_C^3 \rangle & = & \frac{8}{35} \langle  | \mathbf{u} |^6 \rangle  \langle \lambda_1 \lambda_2 \lambda_3 \rangle.
\label{eq:pC_trd_decor_int}
\end{eqnarray}
Using the relation $\langle \lambda_1 \lambda_2 \lambda_3 \rangle = (1/3) \langle \tr( \vec{S}^3 ) \rangle = - (1/4) \langle \vec{\omega} \cdot \vec{S} \cdot \vec{\omega} \rangle$~\cite{Betchov56,Siggia81}, one finally obtains:
\begin{equation}
\langle p_C^3 \rangle 
= \frac{8}{105} \langle  | \mathbf{u} |^6 \rangle  \langle \tr(\vec{S}^3) \rangle 
 =   - \frac{2}{35} \langle  | \mathbf{u} |^6 \rangle  \langle \vec{\omega} \cdot \vec{S} \cdot \vec{\omega} \rangle .
\label{eq:pC_trd_decor_fin}
\end{equation}
Thus, Equation~\eqref{eq:pC_trd_decor_fin} relates the third moment of $p_C$
to vortex stretching 
in such a way that positive vortex stretching
($\langle \vec{\omega} \cdot \vec{S} \cdot \vec{\omega} \rangle > 0$) gives
rise to a negative value of $\langle p_C^3 \rangle$.

Furthermore, many experimental and numerical studies show that 
the probability distributions of individual components of velocity 
$\vec{u}$ are close to Gaussian,
with small deviations that can be quantitatively explained
(see e.g.,~\cite{FL97,WDF11}).
Assuming a Gaussian distribution of $\mathbf{u}$ allows us 
to express the $6^{th}$ moment
of velocity in Eq.~\eqref{eq:pC_trd_decor_fin} in terms of the velocity 
variance $ \langle \mathbf{u}^2 \rangle$. 
Using other known identities,
in particular concerning the relation between $\langle \tr( \vec{S}^3 ) \rangle$
and the skewness of the
velocity derivative $S_{\partial_x u_x}$, as explained in Appendix A,
Eq.~\eqref{eq:pC_trd_decor_fin} can be written as:
\begin{equation}
\langle p_C^3 \rangle = \frac{7}{225} S_{\partial_x u_x} R_\lambda^3 \dissip^3 .
\label{eq:pCcub_Sdudx}
\end{equation} 
The weak dependence of the velocity derivative skewness on the Reynolds 
number, $S_{\partial_x u_x} \propto R_\lambda^{\delta}$~\cite{SA97,Ishihara07}
in Eq.~\eqref{eq:pCcub_Sdudx}
suggests a small correction to the simple order of magnitude analysis for the third moment: 
$\langle p_C^3 \rangle \propto R_\lambda^{3 + \delta}$ 
with $\delta \approx 0.1$.

The assumptions of lack of correlation
between $\vec{u}$ and $\vec{S}$, 
and of a Gaussian distribution of the velocity $\mathbf{u}$,
also lead to an exact determination of the variance of $p_C$:
$ \langle p_C^2 \rangle = \frac{1}{15} R_\lambda^2 \dissip^2 $, see 
Appendix A.
This expression for the second moments of $ p_C$ provides further 
justification for
the dimensional estimate of the variance of $p_C$, and is found to be in 
very good agreement with our DNS results (see Table~\ref{table:mom2}).

Having established the relation between the third moment $\langle p_C^3 \rangle$ 
and vortex stretching, we now establish that the third moment $\langle p^3 \rangle$ is 
dominated by $\langle p_C^3 \rangle$. 
To this end, we first consider
the cancellation between $p_C$ and $p_L$. 

\begin{figure}[h!]
\begin{center}
\includegraphics[width=0.5\textwidth]{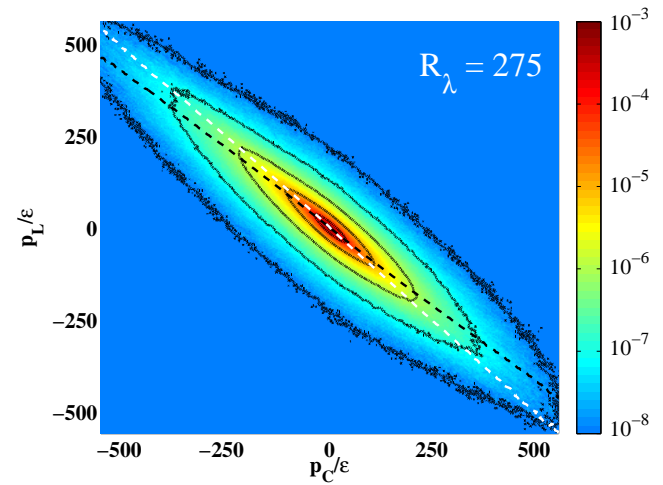}
\end{center}
\caption{The joint probability density function (PDF) between $p_C/\dissip$ (horizontal) 
and $p_L/\dissip$ (vertical) at $R_\lambda = 275$, 
color-coded in a logarithmic scale (see color-bar).
Equal-probability contours, separated by factors of $10$, are shown.
The PDF is concentrated close to the $p_C + p_L = 0$
line, indicating that the two quantities $p_C$ and $p_L$ 
are nearly anti-correlated with each other.
The black dashed line shows $\langle p_L | p_C \rangle / \dissip$, 
which is approximately $-0.86 \times p_C/\dissip$.
The white dashed line shows $\langle p_C | p_L \rangle / \dissip$, 
which is approximately $- p_L/\dissip$.
}
\label{fig:jpdf_pC_pL}
\end{figure}

\begin{figure*}[ht!]
\begin{center}
\subfigure[]{
\includegraphics[width=0.4\textwidth]{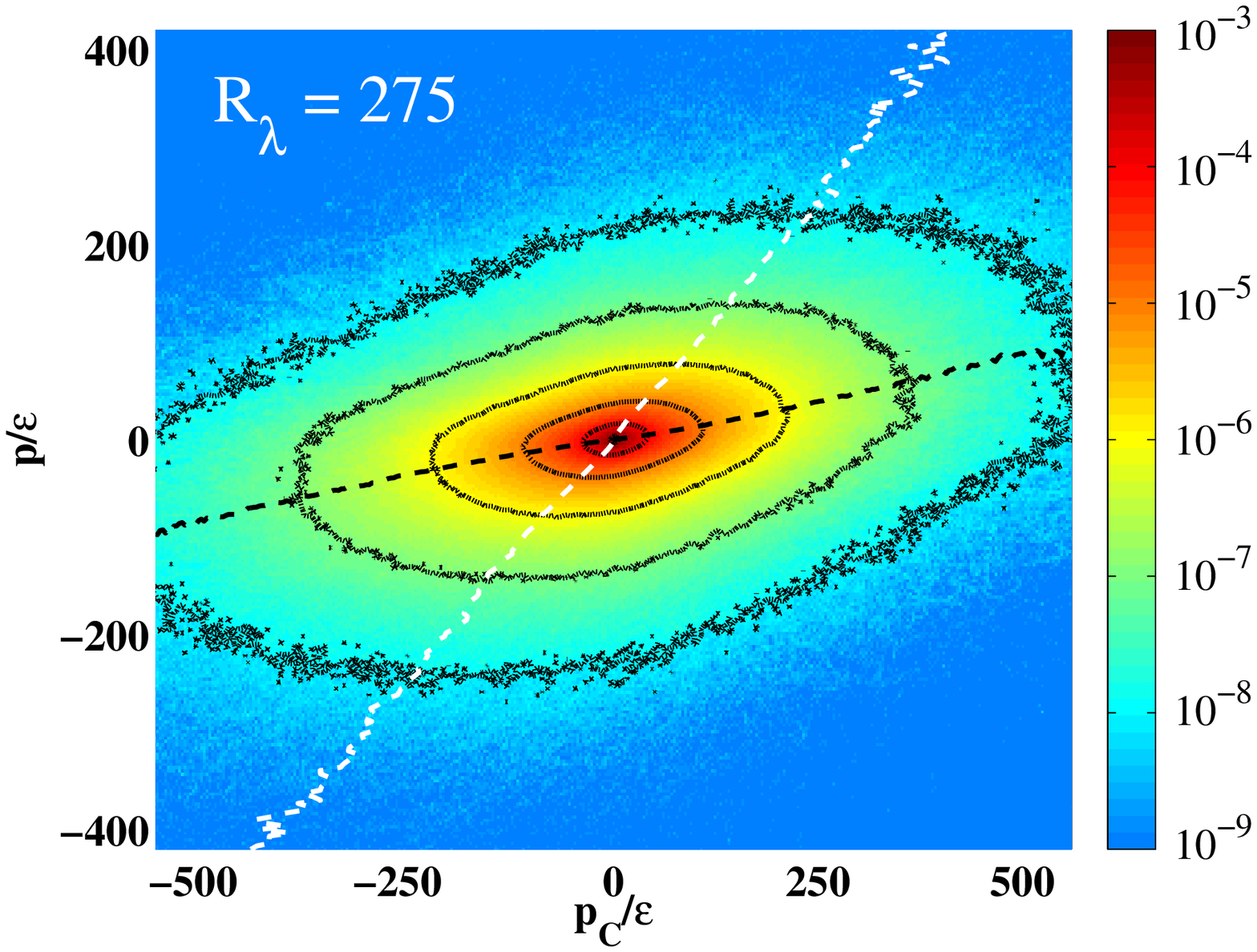}
}
\subfigure[]{
\includegraphics[width=0.4\textwidth]{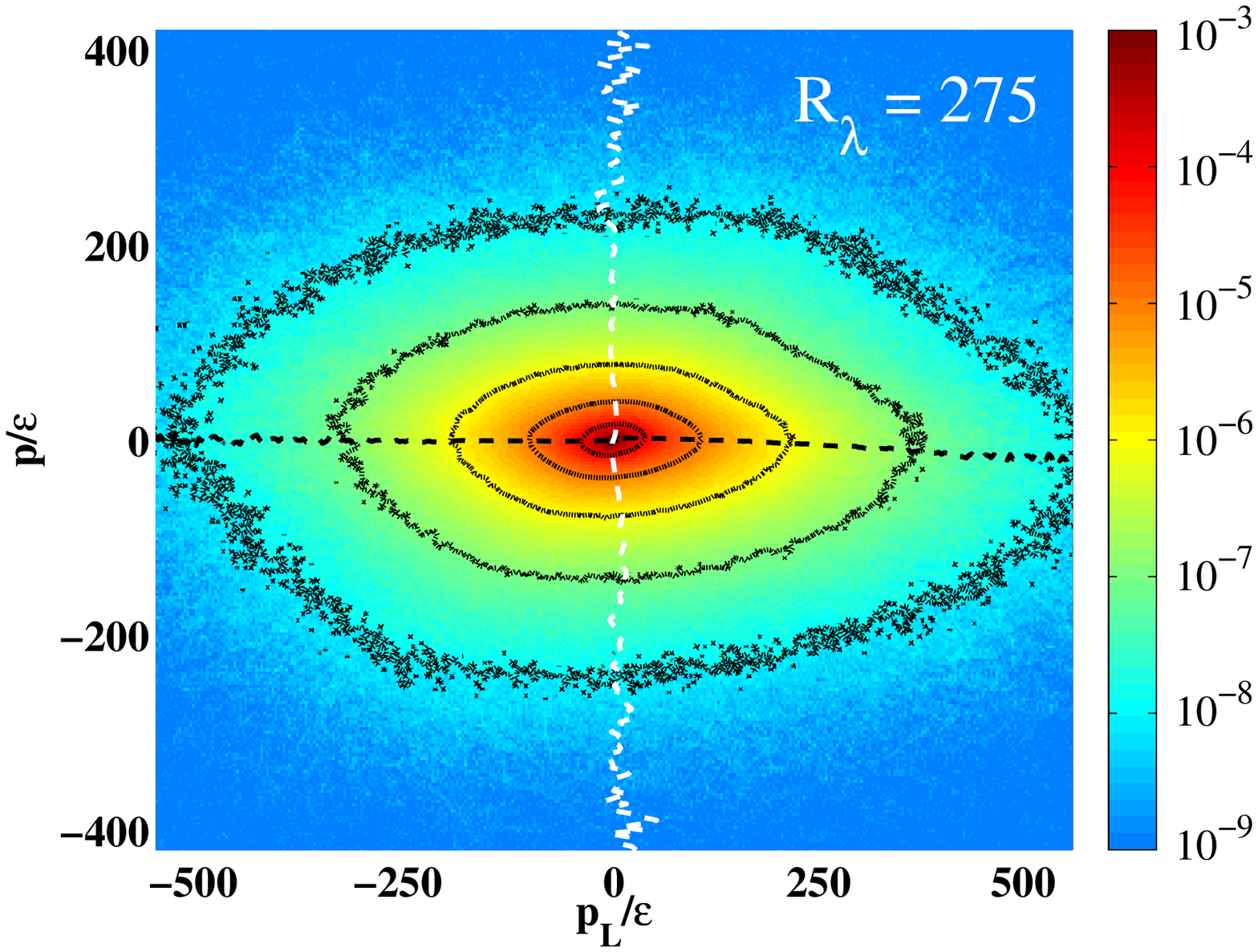}
}
\end{center}
\caption{
(a) The joint PDF between $p_C/\dissip$ (horizontal) and 
$p/\dissip$ (vertical), and (b) The joint PDF between 
$p_L/\dissip$ (horizontal) and $p/\dissip$ (vertical), all at $R_\lambda = 275$,
color-coded in a logarithmic scale (see color-bar).
The equal-probability contours shown are separated by factors of $10$.
Note that the ranges of values of $p/\dissip$ shown are smaller than those of $p_C/\dissip$ and $p_L / \dissip$. 
In (a), the black dashed line represents the conditional average of 
$p/\dissip$ on $p_C/\dissip$, and is very 
close to $ \langle p | p_C \rangle /\dissip \approx 0.14 \times p_C /\dissip$. 
The white dashed line corresponds to the averaged of $p_C/\dissip$,
conditioned on $p/\dissip$, and is well approximated by 
$\langle p_C | p \rangle/\dissip \approx p/\dissip $.
In (b), the black dashed line represents the conditional average of $p/\dissip$ on $p_L/\dissip$, and is very 
close to $ \langle p | p_L \rangle /\dissip \approx 0$.
The white dashed line indicates to the average of $p_L/\dissip$,
conditioned on $p/\dissip$, and is also very well approximated by 
$\langle p_L | p \rangle /\dissip \approx 0 $.
}
\label{fig:jpdf_pC_p}
\end{figure*}

\subsection{Cancellation between $p_L$ and $p_C$}
\label{sub:2ndmoments}

We note that in homogeneous and stationary flows,
the first moments of $p$, $p_C$ and $p_L$ are all exactly $0$.
Table~\ref{table:mom2} shows that the correlation coefficient between 
$p_L$ and $p_C$:
$\langle p_C p_L \rangle/(\langle p_C^2 \rangle \langle p_L^2 \rangle)^{1/2}$,
is approximately $-0.9$ and seems to approach $-1$ as the Reynolds number 
increases.
This strong anti-correlation results in significant cancellation between 
$p_C$ and $p_L$, so the variance of $p$ is much smaller than those 
of $p_C$ and $p_L$.

Although the range of values of $R_\lambda$ covered by the present study
is not sufficient to reach unambiguous conclusions, our results are generally 
consistent with the expected scalings: $\langle p_{L}^2 \rangle \sim \langle p_C^2 \rangle \propto R_\lambda^2$, and $\langle p^2 \rangle \propto R_\lambda^{4/3}$~\cite{XPFB14}.

Further insight into the strong cancellation between 
$p_C$ and $p_L$ can be gained by studying the joint probability 
density function (PDF) of $p_C$ and $p_L$, shown in Fig.~\ref{fig:jpdf_pC_pL}
for our flow at $R_\lambda = 275$.
Fig.~\ref{fig:jpdf_pC_pL} clearly indicates that with a high probability,
the values of
$p_C$ and $p_L$ are concentrated
close to the line $p_C + p_L = 0$, thus implying a significant
cancellation between the two quantities.
The observed tendency of the correlation coefficient between $p_C$ and 
$p_L$ to approach $-1$ as the 
Reynolds number increases implies that the joint PDF of $p_C$ and $p_L$ 
becomes increasingly 
concentrated around the line $p_L + p_C = 0$ at higher Reynolds numbers.
In all our numerical simulations, we find an approximately linear relation between the 
conditional average $\langle p_L | p_C \rangle$ and $p_C$ (shown as the black 
dashed line in Fig.~\ref{fig:jpdf_pC_pL}): 
$\langle p_L | p_C \rangle \approx - \beta(R_\lambda) p_C$, where 
the dimensionless coefficient $\beta(R_\lambda)$ depends weakly on
$R_\lambda$. In agreement with the observed 
tendency of $p_L$ and $p_C$ to become 
increasingly anti-correlated as $R_\lambda$ increases, we find 
that $\beta(R_\lambda)$ 
slightly increases with the Reynolds number, see Table~\ref{table:mom2}.
This implies that
$\langle p | p_C \rangle  \approx (1-\beta) p_C$,
where the coefficient $1 - \beta$ decreases
as $R_\lambda$ increases, from $\approx 0.17$ at $R_\lambda = 193$ 
to $\approx 0.10$ at $R_\lambda = 430$.
We also observe that the 
average of $p_C$ conditioned on $p_L$, shown as the white dashed line
in Fig.~\ref{fig:jpdf_pC_pL} is 
almost exactly equal to $- p_L$, which implies that 
$\langle p | p_L \rangle \approx 0$.

Fig.~\ref{fig:jpdf_pC_p} shows the joint PDFs of $p_C$ and $p$ (a)
and of $p_L$ and $p$ (b). The conditional averages 
$\langle p | p_C \rangle $ and $\langle p | p_L \rangle$ are shown as
black dashed lines, whereas the conditional averages
$\langle p_C | p \rangle $ and $\langle p_L | p \rangle$ are shown  
as white dashed lines.
The conditional averages of $p_C$ and $p_L$ on $p$ have the particularly
simple forms: $\langle p_C | p \rangle \approx p$ and
$\langle p_L | p \rangle \approx 0$.
In addition, the joint PDF of $p$ and $p_L$ is almost symmetrical to both 
$p = 0$ and $p_L = 0$. 
The power $p$ is therefore well correlated with $p_C$, but not with $p_L$.

\subsection{Prevalence of $p_C$ on the moments of $p$}
\label{sec:simple_model}

\subsubsection{General assumptions}

The prevalence of $p_C$ on the statistical properties of $p$
shown by our numerical results leads to the conclusion that the second
and third moment of $p$ are expressible in terms of the corresponding
moments of $p_C$. To justify this claim, we use the two following results.

{\bf A}
The numerical results shown in Fig.~\ref{fig:jpdf_pC_pL} demonstrate that 
at a fixed value of $p_C$, $p_L | p_C$ fluctuates around the mean value 
$ \langle p_L | p_C \rangle \approx -\beta p_C$.  
This immediately implies the following relations:
\begin{equation}
\langle p_L p_C \rangle = -\beta \langle p_C^2 \rangle \mbox{ and } \langle p_L p_C^2 \rangle = -\beta \langle p_C^3 \rangle ,
\label{eq:pLpC^n-1}
\end{equation}
which can be easily justified by writing $p_L $ conditioned on a value
of $p_C$ as:
\begin{equation}
p_L | p_C = - \beta p_C + \xi | p_C , 
\label{eq:pLmodel}
\end{equation}
where $\xi | p_C$ is a random variable with zero mean and its distribution depends on $p_C$. 
Eq.~\eqref{eq:pLpC^n-1} is found to be numerically extremely well satisfied, see Table~\ref{table:mom3},
as a direct consequence of the quality of the linear dependence between
$\langle p_L | p_C \rangle$ and $p_C$.

{\bf B}
The lack of correlation between $p$ on $p_L$, demonstrated in 
Fig.~\ref{fig:jpdf_pC_p} and manifested by the two relations
$\langle p_L | p \rangle \approx 0$ and $\langle p | p_L \rangle \approx 0$,
implies that:
\begin{equation}
\langle p p_L \rangle \approx \langle p^2 p_L \rangle \approx \langle p p_L^2 \rangle \approx 0 .
\label{eq:constraint_p_pL}
\end{equation}
The equalities shown in Eq.~\eqref{eq:constraint_p_pL} are only approximate.
In the following, we explore
the consequences of the independence between $p$ and $p_L$ by assuming for now
that these equalities are
exactly satisfied, leaving for later a discussion of the errors made.

As we show below the approximations {\bf A} and {\bf B}
above lead to 
a very accurate prediction of all the second moments of $p_C$, $p_L$ and $p$,
in terms of $\langle p_C^2 \rangle$ and $\beta$. 
The predictions concerning the third moments, 
however, are not as accurate as those concerning the second moments, 
as a consequence of quantitative deviations from the symmetry 
assumption {\bf B}. 

\subsubsection{Second moments}
\label{subsec:2nd_mom}

Using Eq.~\eqref{eq:pLpC^n-1},
$\langle p_L p_C \rangle = - \beta \langle p_C^2 \rangle$,
and Eq.~\eqref{eq:constraint_p_pL}, $\langle p p_L \rangle = 0$,
we determine the second moment of $p_L$ as a function of 
$\langle p_C^2 \rangle$:
$\langle p_L^2 \rangle = \beta \langle p_C^2 \rangle$, from which we obtain:
\begin{equation}
\langle p^2 \rangle = (1-\beta) \langle p_C^2 \rangle .
\label{eq:pvariance2}
\end{equation}
We find that the condition of independence $\langle p p_L \rangle = 0$ is
very well satisfied, which implies that
Eq.~\eqref{eq:pvariance2} is numerically very accurately satisfied
(see Appendix B and Table~\ref{table:AppendB}).

\subsubsection{Third moments}
\label{subsec:3rd_mom}

The results from Eq.~\eqref{eq:pCcub_Sdudx}, showing that 
$\langle p_C^3 \rangle \approx - \dissip^3 R_\lambda^3 $,
together with the observation that 
$\langle p^3 \rangle \approx - \dissip^2 R_\lambda^2$ \cite{XPFB14}, 
also point to a strong cancellation
between $p_C $ and $p_L$ in the third moment $\langle p^3 \rangle$. 
To relate the properties of the third moments of $p$ to 
$\langle p_C^3 \rangle$,
we begin by noting that Eq.~\eqref{eq:constraint_p_pL} 
leads to the following expressions for the third order moments: 
$ \langle p_L^3 \rangle = - \beta \langle p_C^3 \rangle$ and
$\langle p_L^2 p_C \rangle = \beta \langle p_C^3 \rangle$, and hence
to the expression $\langle p^3 \rangle = ( 1 - \beta ) \langle p_C^3 \rangle$.
These expressions predict simple relations between the various
moments $\langle p_C^m p_L^n \rangle$ with $m + n = 3$ and 
$\langle p_C^3 \rangle$, and lead to the correct sign of $\langle p^3 \rangle$,
thus justifying our claim that the assumption of independence beween $p$ and 
$p_L$ imposes that the sign of $\langle p^3 \rangle $ is given by $\langle p_C^3 \rangle$.

The expressions obtained above, however, are quantitatively
not accurate. The reason is that while $\langle p_L p^2 \rangle$
is found to be very small (of the order of 
$1 \%$ of $| \langle p_C^3 \rangle |$),
the numerical values of $\langle p_L^2 p \rangle$ are found to be much 
larger, of the order of $10 \%$ of $ | \langle p_C^3 \rangle |$. 
The small, but significant error in $\langle p_L^2 p \rangle = 0$ therefore
leads to a significant reduction of the numerical value of 
$\langle p^3 \rangle$, consistent with the numerical values shown in
Table~\ref{table:mom3}.
To take the effect of non-zero $\langle p_L^2 p \rangle$
into account, we denote 
$\zeta =  \langle p_L^2 p \rangle / \langle p_C^3 \rangle$, 
where $\zeta$ is a positive number of order 
$\sim 0.1$ (see Table~\ref{table:mom3}).
and decreases when $R_\lambda$ increases.
This then leads to 
$\langle p_L^2 p_C \rangle = (\beta - \zeta ) \langle p_C^3 \rangle$
and
$\langle p_L^3 \rangle = -(\beta - 2 \zeta ) \langle p_C^3 \rangle$, 
and consequently:
\begin{equation}
\langle p^3 \rangle = ( 1 - \beta - \zeta) \langle p_C^3 \rangle .
\label{eq:thrd} 
\end{equation}

Using Eq.~\eqref{eq:thrd} and the relation between $\langle p_C^3 \rangle$
and vortex stretching, Eq.~\eqref{eq:pC_trd_decor_fin}, we obtain:
\begin{equation}
\langle p^3 \rangle =  -\frac{2}{35} (1-\beta - \zeta) \langle  | \mathbf{u} |^6 \rangle  \langle \vec{\omega} \cdot \vec{S} \cdot \vec{\omega} \rangle ,
\end{equation}
which establishes a quantitative relation between the time irreversibility, as measured by $\langle p^3 \rangle$, and vortex stretching, a small-scale generation mechanism in 3D turbulence.

We note that Eq.~\eqref{eq:pvariance2}, together with the observed
scaling $\langle p^2 \rangle \propto R_\lambda^{4/3}$ and 
$\langle p_C^2 \rangle \propto R_\lambda^{2}$, suggests that
$( 1 - \beta) \propto R_\lambda^{-2/3}$. Similarly, the dependence
$\langle p_C^3 \rangle \propto R_\lambda^3$, 
together with the observation of \cite{XPFB14} that $\langle p^3 \rangle
\approx - \dissip^3 R_\lambda^2$, imply, using
Eq.~\eqref{eq:thrd}, that $(1 - \beta - \zeta) \propto R_\lambda^{-1}$
or $1 - \zeta/(1-\beta) \propto R_\lambda^{-1/3}$.
These are consistent with the values obtained numerically.
As shown in Table~\ref{table:mom3}, the value of $1 - \zeta/(1-\beta)$ decreases
slightly, from $0.36$ to $0.34$, when the Reynolds number increases from $R_\lambda = 193$ to $430$.

The results presented here thus show that, while the statement of independence 
between $p$ and $p_L$ is merely an enticing approximation, taking quantitatively
into account the deviations from Eq.~\eqref{eq:constraint_p_pL} does not affect
our main conclusion: the third moment of $p$ is controlled by 
the third moment of $p_C$.

%%%%%%%%%%%%%%%%%%%%%%%%%%%%%%%%%%%%%%%%%%%%%%%%%%%%%%%

\begin{figure}[h!]
\includegraphics[width=0.5\textwidth]{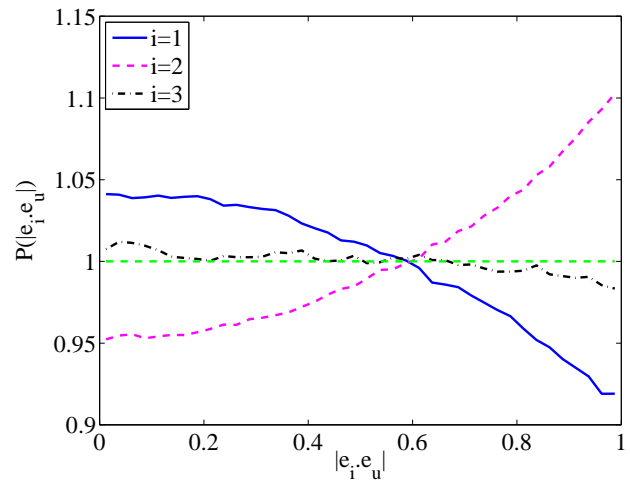}
\caption{The PDFs of the cosines of the angles between the direction of the velocity $\mathbf{u}$, and
the eigenvectors $\mathbf{e}_i$ of $\mathbf{S}$ at $R_\lambda = 275$. 
The cosine of the angles is given by the inner product of the two unit vectors: $\hat{x}_i = \vec{e}_i \cdot \vec{e}_u$, 
where $\vec{e}_u = \vec{u} / | \vec{u} |$.
A lack of correlation between $\mathbf{u}$ and $\mathbf{S}$ would lead to
a flat PDF of $\vec{e}_i \cdot \vec{e}_u$: ${\cal P}(| \mathbf{e}_i \cdot \mathbf{e}_u |) = 1$, which
are supported by the data to a good degree. The small
departure from this expectation points to a weak correlation 
between $\mathbf{u}$ and $\mathbf{S}$.
}
\label{fig:PDF_ue_i_incond}
\end{figure}

\begin{figure}[h!]
\centering
\includegraphics[width=0.45\textwidth]{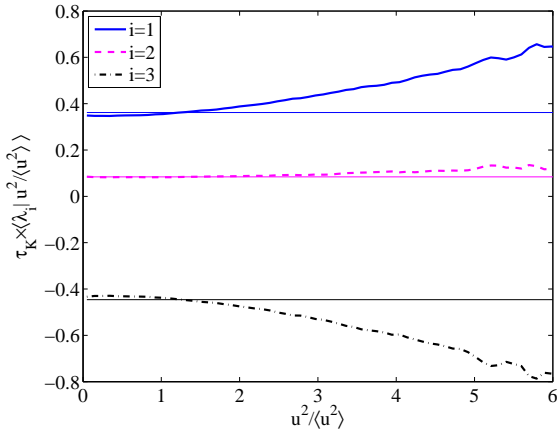}
\caption{The thick lines show the conditional averages of the eigenvalues of $\mathbf{S}$, 
$\langle \lambda_i | \vec{u}^2 \rangle$, made dimensionless
by $\tau_K = 1/ (2 \langle \vec{S}^2 \rangle)^{1/2}  = 1/ (2 \sum \lambda_i^2)$.
The thin lines with the same color are the corresponding unconditional averages $\langle \lambda_i \rangle$, all at $R_\lambda = 275$.
For small values of $\vec{u}^2$, the conditional averages are nearly the same as the corresponding
unconditional averages, consistent with the assumption that $\vec{u}$ and $\mathbf{S}$ are uncorrelated.
For large $\vec{u}^2$, the magnitudes of the conditional averages increase with $\vec{u}^2$, 
indicating a deviation from the assumption.
Note that because the probability of having large values of $\mathbf{u}^2$ decreases very rapidly with $\mathbf{u}^2$, the increases of $\langle \lambda_i | \mathbf{u}^2 \rangle$ at large $\mathbf{u}^2$ have only very small effects on the unconditional averages $\lambda_i$.
}
\label{fig:cond_S_u2}
\end{figure}

\subsection{Lack of correlation between $\mathbf{u}$ and $\mathbf{S}$}
\label{subsec:correl_S_u}

We return here briefly to discuss the essential assumption that
$\mathbf{u}$ and $\mathbf{S}$ are uncorrelated. 
Specifically, we examine in this subsection
the correlation between the angles of 
$\vec{u}$ and the eigenvectors $\vec{e}_i$ and between the magnitude of 
$\vec{u}$ and the eigenvalues $\lambda_i$.
In the following, the values $\lambda_i$ are sorted in decreasing 
order: $\lambda_1 \ge \lambda_2 \ge \lambda_3$.

Fig.~\ref{fig:PDF_ue_i_incond} shows the PDFs of
$|\hat{x}_i |= | \mathbf{e}_i \cdot \mathbf{e}_u |$, 
the absolute value of the cosine of the angle between
the eigenvector $\mathbf{e}_i$ and the unit vector in the direction 
of the velocity $\mathbf{e}_u = \vec{u} / |\vec{u}|$
(the sign of this cosine is immaterial) at $R_\lambda = 275$.
A complete lack of correlation between $\mathbf{u}$ and $\mathbf{S}$ 
implies that the PDFs of $| \mathbf{e}_i \cdot \mathbf{e}_u |$ are  
constant and equal to $1$. Fig.~\ref{fig:PDF_ue_i_incond} shows that 
this is close to be true. 
Namely, the probability of alignment between $\mathbf{e}_1$ and 
$\vec{u}$, i.e., of 
$| \mathbf{e}_u \cdot \mathbf{e}_1 | $ being close to $1$, is 
slightly reduced.
On the contrary, the probability of alignment between
$\mathbf{e}_u$ and $\mathbf{e}_2$ is slightly increased.
The deviations observed numerically are weak, 
less than $\sim 10 \%$, compared to the uniform distribution. 
The cosine between 
$\mathbf{e}_u$ and $\mathbf{e}_3$, is very close to being 
uniformly distributed.
The nearly uniform PDF of 
$| \mathbf{e}_i \cdot \mathbf{e}_u |$ indicate
that the assumption that $\mathbf{S}$ and $\mathbf{u}$ are uncorrelated,
explicitly used in the determination of $\langle p_C^3 \rangle$, 
provides a very good first order approximation.

The assumption that $\vec{u}$ and $\vec{S}$ are uncorrelated also implies that the
conditional averages of the properties of $\vec{S}$ should be independent 
of the magnitude of $\vec{u}$.

Fig.~\ref{fig:cond_S_u2} shows that the dependence of the 
conditional average of the eigenvalues of $\vec{S}$
on $ \vec{u}^2$, 
$\langle \lambda_i | \mathbf{u}^2 \rangle$,  
is weak.
Systematic deviations are visible at large values of $\vec{u}^2 $, 
where the magnitudes of the averaged conditional eigenvalues are larger.
The probability of large values of $\vec{u}^2$, however,
drops very rapidly when $\vec{u}^2$ increases~\cite{FL97,GFN02,WDF11},
so the effect of this weak dependence
of $\lambda_i$ on $\vec{u}^2$ has only a small effect on the 
low-order moments of $p_C$ studied here.

In summary, 
the results presented here and in the Appendix C show that
the assumption of a lack of correlation between $\vec{u}$ and $\vec{S}$ 
provides a very good first-order approximation 
to describe the third moment of $p_C$.

%%%%%%%%%%%%%%%%%%%%%%%%%%%%%%%%%%%%%%%%%%%%%%%%%%%%%%%

\section{Discussion and Conclusion}
\label{sec:conclusion}

Our work, aimed at understanding the third moment of
the power $p$ acting on fluid particles $\langle p^3 \rangle \approx -
\dissip^3 R_\lambda^2 $, and
its implication for the physics of turbulent flows~\cite{XPFB14}, 
rests on decomposing $p$ into two parts: 
a local part, $p_L = \vec{u} \cdot \partial_t \vec{u}$,
induced by the change of the kinetic energy at a fixed spatial point,
and a convective part, $p_C = \vec{u} \cdot \nabla (\vec{u}^2 / 2)$, 
due to 
the change in kinetic energy along particle trajectories, assuming the velocity field
is frozen.
We observe that the two terms $p_C$ and $p_L$ cancel each other to a large extent, 
resulting in a much smaller variance of $p$ compared to those of 
either $p_C$ or $p_L$.
This cancellation may be qualitatively explained by invoking a fast sweeping
of the small scales of the flow by the large scales~\cite{Tennekes71}.
In physical words, 
kinetic energy along particle trajectories, is mostly carried (swept) by the 
flow, and 
changes far less than it would change by keeping the flow fixed, or by varying
the flow with the same position in time. This fact has been documented in
a slightly different context~\cite{Kamps09}.
Our results provide
a quantitative characterization of how much sweeping reduce the individual 
contributions of $p_L$ and $p_C$.

One of the two main results of our work is that the third moment of $p_C$, 
expressed in terms of the rate of strain, $\vec{S}$,
and the velocity, $\vec{u}$, as:
$p_C = \mathbf{u} \cdot \mathbf{S} \cdot \mathbf{u}$, can be exactly 
determined, by using the physically justified approximation that 
$\mathbf{u}$ and 
$\mathbf{S}$ are uncorrelated.
Remarkably, we find that 
$\langle p_C^3 \rangle$ is directly related to vortex stretching,
$\langle \mathbf{\omega} \cdot \mathbf{S} \cdot \mathbf{\omega} \rangle$.
In particular,
the {\it negative} sign of $\langle p_C^3 \rangle$ originates from the 
{\it positive} sign of the vortex stretching, 
due to small-scale generation by turbulence.
This observation provides the first basis 
for our claim that the third moment of $p$ is related to the generation 
of small scales
in 3D turbulent flows.

The other main observation of our work is that, despite  
the strong cancellation between $p_L$ and $p_C$, the
power $p$ correlates with $p_C$, but not with $p_L$,
as revealed by the nearly vanishing conditional averages $\langle p | p_L \rangle \approx 0$ and $\langle p_L | p \rangle \approx 0$.
Assuming these conditional averages are exactly zero leads to a simple relation between $\langle p^3 \rangle$ and $\langle p_C^3 \rangle$.
The (weak) corrections to this simple assumption modify 
only quantitatively the results. 

Taken together, these two observations, namely that
$\langle p^3 \rangle$ is controlled by $\langle p_C^3 \rangle$ 
and that $\langle p_C^3 \rangle$ is directly linked to vortex stretching, 
$\langle \mathbf{\omega} \cdot \mathbf{S} \cdot \mathbf{\omega} \rangle > 0$, 
allow us to establish a relation
between the third moment of power, $\langle p^3 \rangle$, and  
vortex stretching.
Thus, the recently observed manifestation of irreversibility
in studying the statistics of individual Lagrangian trajectories can be
understood as resulting from small-scale generation in 3D 
turbulent flow.

For lack of essential information concerning the quantities investigated here, 
our work rests on several assumptions supported by numerical observations.
The well-known fact that 
velocity, $\vec{u}$, and the rate of strain, $\vec{S}$, 
are dominated by large- and small-scales, respectively, 
makes it plausible that these two quantities are mostly uncorrelated. 
Our numerical results
confirm this expectation. Although small, and of little relevance for 
the low-order moments studied here, the deviations observed 
suggest an interesting structure, which would be worth elucidating.
The observation that $p$ and $p_L$ are not correlated, in the 
sense that the conditional averages $\langle p | p_L \rangle $ and 
$\langle p_L | p \rangle$ are both very close to zero, rests only on
numerical observations, and requires a proper explanation.
Understanding and quantifying the weakness of the correlation between 
$p$ and $p_L$ in 3D turbulent flows may provide
important hints not only on higher moments of $p$, but more importantly,
on the structure of the flow itself.

We note that studying the cancellation between $p_C$ and $p_L$ by
directly focusing on the effect of the pressure gradient, $ - \vec{u}\cdot \nabla P$ is
likely to 
lead to satisfactory results when studying the second moments of $p$,
as the pressure term has been documented 
to provide the largest contribution to the variance 
$\langle p^2 \rangle$~\cite{PXB+14}. In 3D, however, the third moment 
$-\langle ( \mathbf{u} \cdot \nabla P)^3 \rangle $ has been shown
to contribute negligibly to $\langle p^3 \rangle$, whose understanding
requires the investigation of other correlations~\cite{PXB+14}.

Finally, the arguments provided here to explain the negative third
moments of power fluctuations of particles in 3D turbulent flows
should {\it not} be applied to 2D turbulence, in which 
the third moment of $p$ is also negative, and grows with 
a similar power of the Reynolds number~\cite{XPFB14},
but the amplification of large velocity 
gradients is due to entirely different physical processes~\cite{Boffetta:12}.
Still, one may expect that the manifestations of irreversibility, in 2D
turbulence as well as in a broad class of non-equilibrium systems, to be 
fundamentally related to a flux in the system.

\section*{Acknowledgments}
We thank G. Falkovich for insightful comments.
This work is supported by the Max-Planck Society. AP acknowledges
partial support from
ANR (contract TEC 2), the Humboldt foundation, and the PSMN at the
Ecole Normale Sup\'erieure de Lyon.
RG acknowledges the support from the German Research Foundation (DFG) 
through the program FOR 1048.

\section*{Appendix A: Alternative expressions of the third moment of $p_C$ }
\label{sec:alternative}

\begin{table}[t!]
\begin{center}
\begin{tabular}{|c||c|c|c|}
\hline
$R_{\lambda}$ & $193$ & $275$ & $430$ \\ 
\hline
$ 225 \langle p_C^3 \rangle/(7 \dissip^3 R_\lambda^3)$ & $ -0.24 $ & $-0.37$ & $-0.40$ \\
%$ \frac{\sqrt{15}}{7} \times S_{p_C} \equiv \langle p_C^3 \rangle / \langle p_C^2 \rangle^{3/2}$ & $-0.30$ & $-0.38$ & $-0.42$ \\  
$ \frac{\sqrt{15}}{7} \times S_{p_C}$ & $-0.30$ & $-0.38$ & $-0.42$ \\  
\hline
$S_{p} \equiv \langle p^3 \rangle / \langle p^2 \rangle^{3/2}$ & $-0.52$ & $-0.62$ & $-0.67$ \\  
\hline
\end{tabular}
\caption{
Third moments of the distributions of 
$p_C/\dissip$ and $p/\dissip$ 
at the three Reynolds numbers studied in this article. 
}
\label{table:AppendA}
\end{center}
\end{table}

In many experiments and numerical simulations, 
the probability distributions of individual components of velocity 
$\vec{u}$ are close to Gaussian~\cite{WDF11}. That observation allows us to estimate
explicitly the $6^{th}$ moment of $| \vec{u}|$ in terms of the second moment:
\begin{equation}
\langle | \vec{u} |^4 \rangle = \frac{5}{3} \langle | \vec{u}|^2 \rangle^2 ~ ,
~~ \langle | \vec{u} |^6 \rangle = \frac{35}{9} \langle | \vec{u}|^2 \rangle^3 ,
\label{eq:6th_mom}
\end{equation}
which, when substituted into Eq.~\eqref{eq:pC_trd_decor_fin}, gives an expression for $\langle p_C^3 \rangle$ as:
\begin{equation}
\langle p_C^3 \rangle = \frac{8}{27}
\langle | \vec{u} |^2 \rangle^3 \langle \tr(\mathbf{S}^3 ) \rangle .
\label{eq:pC_u2}
\end{equation}

The same assumptions and similar elementary algebraic manipulations as
those used to establish Eq.~(\ref{eq:pC_u2}), lead to an exact expression for the second moment of $p_C$:
\begin{equation}
\langle p_C^2 \rangle = \frac{2}{15} \langle  | \mathbf{u} |^4 \rangle  \langle \tr(\vec{S}^2) \rangle = \frac{1}{15} R_\lambda^2 \dissip^2 ,
\label{eq:pCvariance_trS}
\end{equation}
This relation is found to be in very good agreement with our numerical
results, see Table~\ref{table:mom2}. 

The known relation between $\tr(\mathbf{S}^3)$ and the experimentally accessible
moments of $\partial_x u_x$~\cite{Betchov56}:
\begin{equation}
\langle \tr( \mathbf{S}^3 ) \rangle = \frac{105}{8} \langle ( \partial_x u_x)^3 \rangle ,
\label{eq:dudx_3rd}
\end{equation}
together with the expression for the second moment of $\partial_x u_x $:
$\langle (\partial_x u_x)^2 \rangle = \dissip/( 15 \nu)$~\cite{Tennekes,frisch95},
leads to the following expression for the third moment of $p_C$:
\begin{equation}
\langle p_C^3 \rangle = \frac{7}{225} S_{\partial_x u_x} R_\lambda^3 \dissip^3 ,
\label{eq:pc3_vs_Sk}
\end{equation}
where 
$S_{\partial_x u_x} \equiv \langle (\partial_x u_x)^3 \rangle/\langle (\partial_x u_x)^2 \rangle^{3/2}$ 
is the skewness of the velocity derivative.

Combining Eq.~(\ref{eq:pCvariance_trS}) and (\ref{eq:pc3_vs_Sk})
gives the following relation between the skewness 
of $\partial_x u_x$ and the skewness of $p_C$:
\begin{equation}
S_{\partial_x u_x} = \frac{\sqrt{15}}{7} S_{p_C}.
\label{eq:skew_pc_S}
\end{equation}
The values of $ \frac{\sqrt{15}}{7} S_{p_C}$, as well as the ratio 
$225 \langle p_C^3 \rangle/ (7 \dissip^3 R_\lambda^3 ) $,
determined from our numerical simulations, are shown in 
Table~\ref{table:AppendA}. The correponding value of the skewness of 
$\partial_x u_x$, $S_{\partial_x u_x }$ is found to be approximately 
$ \approx -0.4$, which is well within the range of values of 
$S_{\partial_x u_x}$ 
reported from experiments and simulations at comparable Reynolds 
numbers~\cite{SA97,Ishihara07}. 

\section*{Appendix B: Prevalence of $p_C$ on the moments of $p$}
\label{sec:prevalence}

\begin{table}[t!]
\begin{center}
\begin{tabular}{|c||c|c|c|}
\hline
$R_{\lambda}$ & $193$ & $275$ & $430$ \\ 
\hline
$\langle p_L ~ p_C \rangle / \langle p_L^2 \rangle^{1/2} \langle p_C^2 \rangle^{1/2}$ & $-0.907$ & $-0.923$ & $-0.944$ \\  
$-\beta^{1/2}$ & $-0.912$ & $-0.928$ & $-0.947$ \\  
\hline
$\langle p^2 \rangle / \langle p_C^2 \rangle$ & $0.178$ & $0.147$ & $0.110$ \\  
$1-\beta$ & $0.17$ & $0.14$ & $0.10$ \\  
\hline
$\langle p_L^2 \rangle / \langle p_C^2 \rangle$ & $0.823$ & $0.851$ & $0.892$ \\  
$\beta$ & $0.83$ & $0.86$ & $0.90$ \\  
\hline
\end{tabular}
\caption{
Parametrisation of the second moments of 
$p_C/\dissip$, $p_L/\dissip$ and $p/\dissip$, compared to the expression in 
terms of $\beta$ given by Eq.~\ref{eq:pL^2pC^2},\ref{eq:pLpC} and \ref{eq:p^2pC^2},
at the three Reynolds numbers studied in this article. 
}
\label{table:AppendB}
\end{center}
\end{table}

\subsection{Second moments of $p$}
\label{subsec:sec_p}

The decomposition of the distribution of $p_L$ conditioned on $p_C$, see Eq.~\eqref{eq:pLmodel}, involving a random variable $\xi |p_C$ with a zero mean,
together with the assumption of independence between $p$ and $p_L$, 
$\langle p p_L \rangle = 0$ (see Eq.~(\ref{eq:constraint_p_pL})),
leads to a full description of the second moments
of $p_C$, $p_L$ in terms of $\beta$ only:
\begin{eqnarray}
& \langle  p_L^2 \rangle & = \beta \langle p_C^2 \rangle \label{eq:pL^2pC^2} \\
& \langle  p_L p_C \rangle & =  - \beta \langle p_{C}^2 \rangle = - \beta^{1/2} ( \langle p_C^2 \rangle \langle p_L^2 \rangle)^{1/2}  \label{eq:pLpC} \\
&\langle  p^2 \rangle  & = (1 -\beta) \langle p_C^2 \rangle \label{eq:p^2pC^2} 
\end{eqnarray}
The values of $\beta$ are measured from the conditional average 
$\langle p_L | p_C \rangle = - \beta p_C$, Eq.~\eqref{eq:pLmodel}.
This allows us to check how accurately are 
Eqs.~\eqref{eq:pL^2pC^2} to \eqref{eq:p^2pC^2} satisfied.
The numerical results are shown in Table~\ref{table:AppendB}.
The relations
given by Eqs.~\eqref{eq:pL^2pC^2} to \eqref{eq:p^2pC^2} are found to
be very well satisfied, with very small errors, thus demonstrating that
the proposed parametrization in
terms of $\langle p_C^2 \rangle $ and $\beta$ provides a very good description
of the second moments of $p$.

\begin{figure*}
\centering
\subfigure[]{
\includegraphics[width=0.48\textwidth]{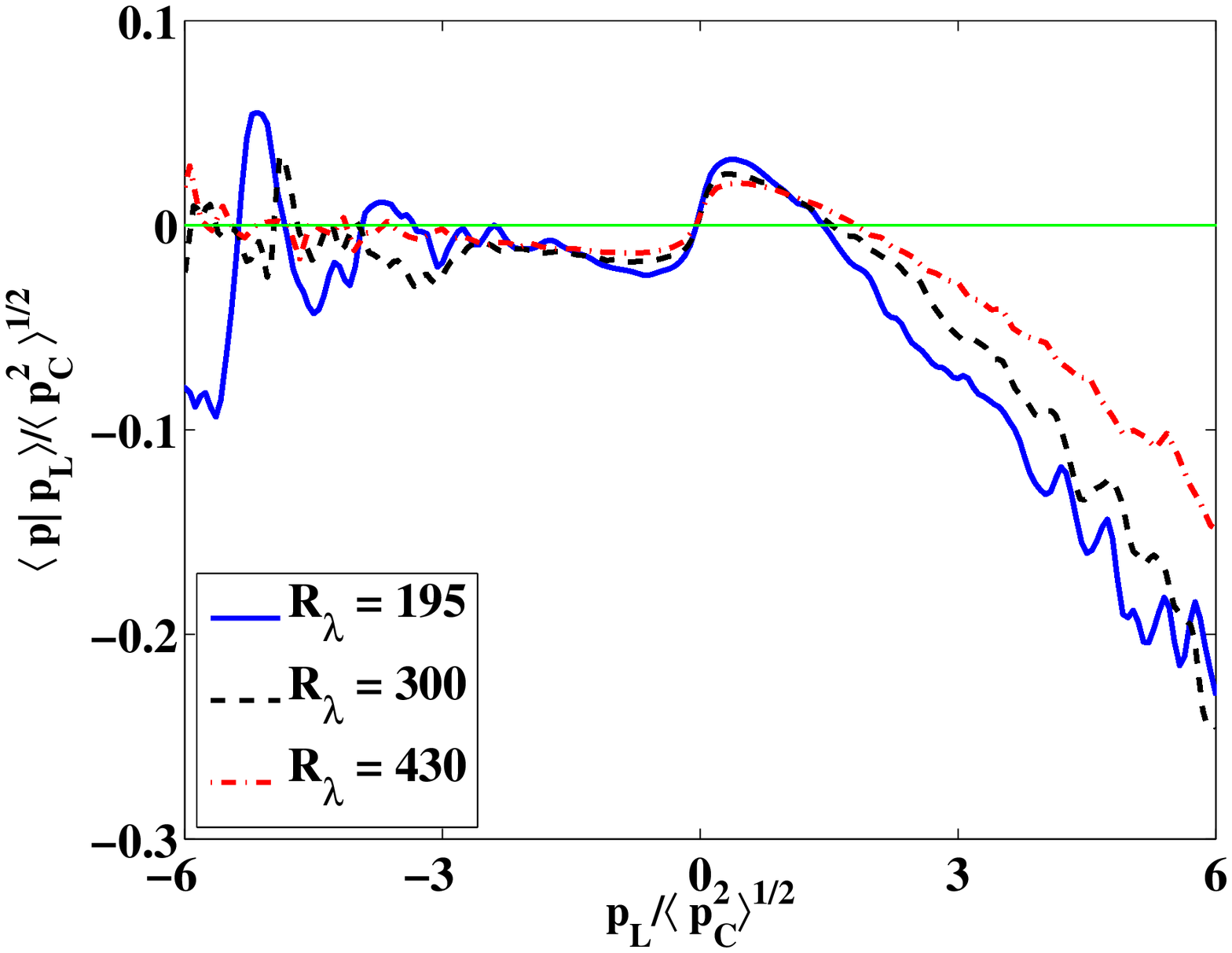}
}
\subfigure[]{
\includegraphics[width=0.48\textwidth]{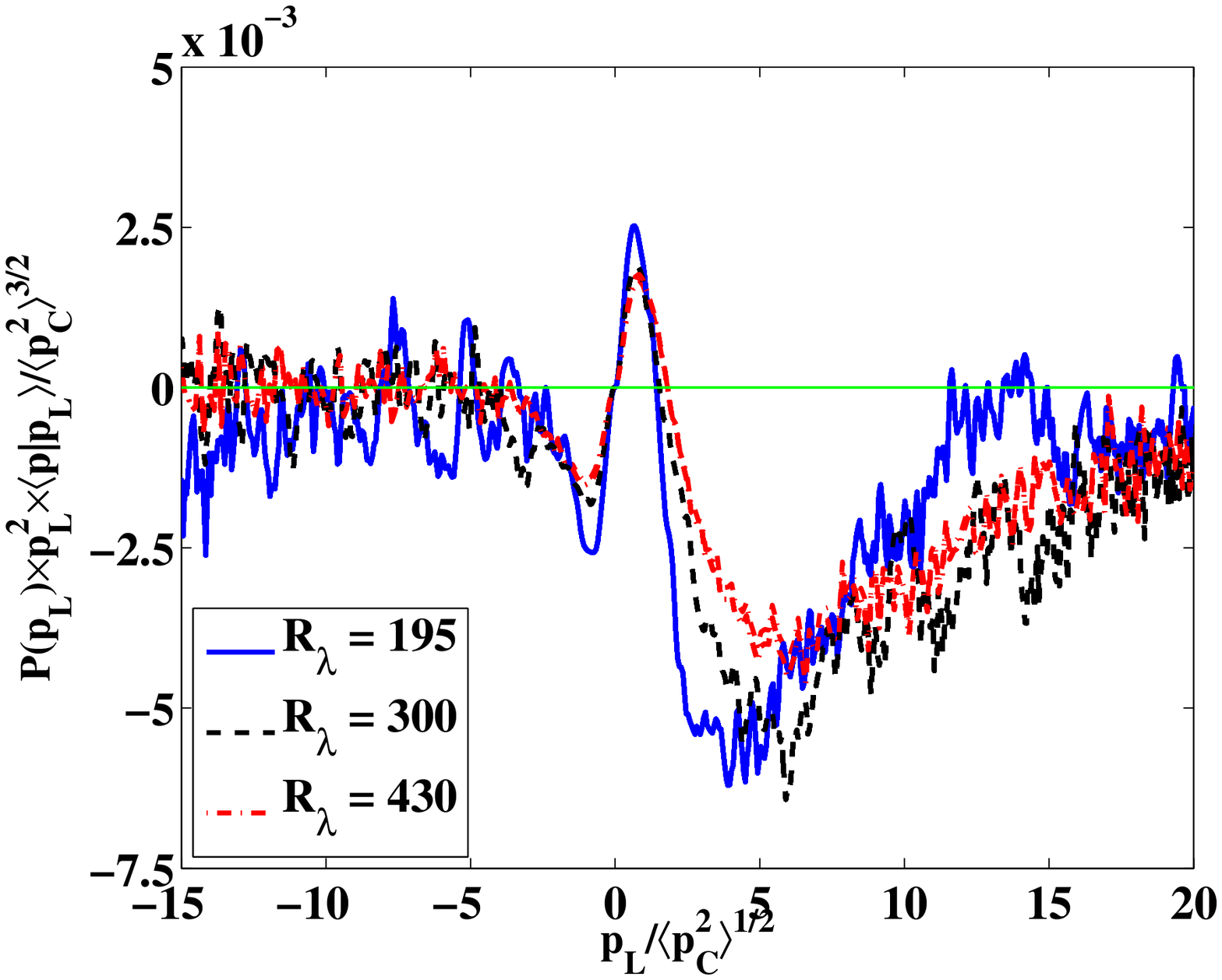}
}
\caption{The conditional average of $p$ on $p_L$: (a) raw data
$\langle p | p_L \rangle / \langle p_C^2 \rangle^{1/2}$ vs. $p_L / \langle p_C^2 \rangle^{1/2}$
and (b) $\mathcal{P}(p_L) \langle p | p_L \rangle p_L^2 / \langle p_C^2 \rangle^{3/2}$,
which provides the integrand in Eq.~(\ref{eq:identity})
to calculate the normalized third moment $\langle p_L^2 p \rangle / \langle p_C^2 \rangle^{3/2} = \zeta S_{p_C}$. 
The conditional average $\langle p | p_L \rangle$ differs weakly, but 
consistently from being $0$, especially for $p_L > 0$, where the
conditional average $\langle p | p_L \rangle$ decreases 
approximately linearly. 
This leads to an appreciable {\it negative} contribution 
to $\langle p_L^2 p \rangle$: $\zeta = \langle p_L^2 p \rangle / \langle p_C^3 \rangle \lesssim 10$\% in the range of Reynolds number studied.
}
\label{fig:cond_p_pL}
\end{figure*}

\begin{figure*}
\centering
\subfigure[]{
\includegraphics[width=0.48\textwidth]{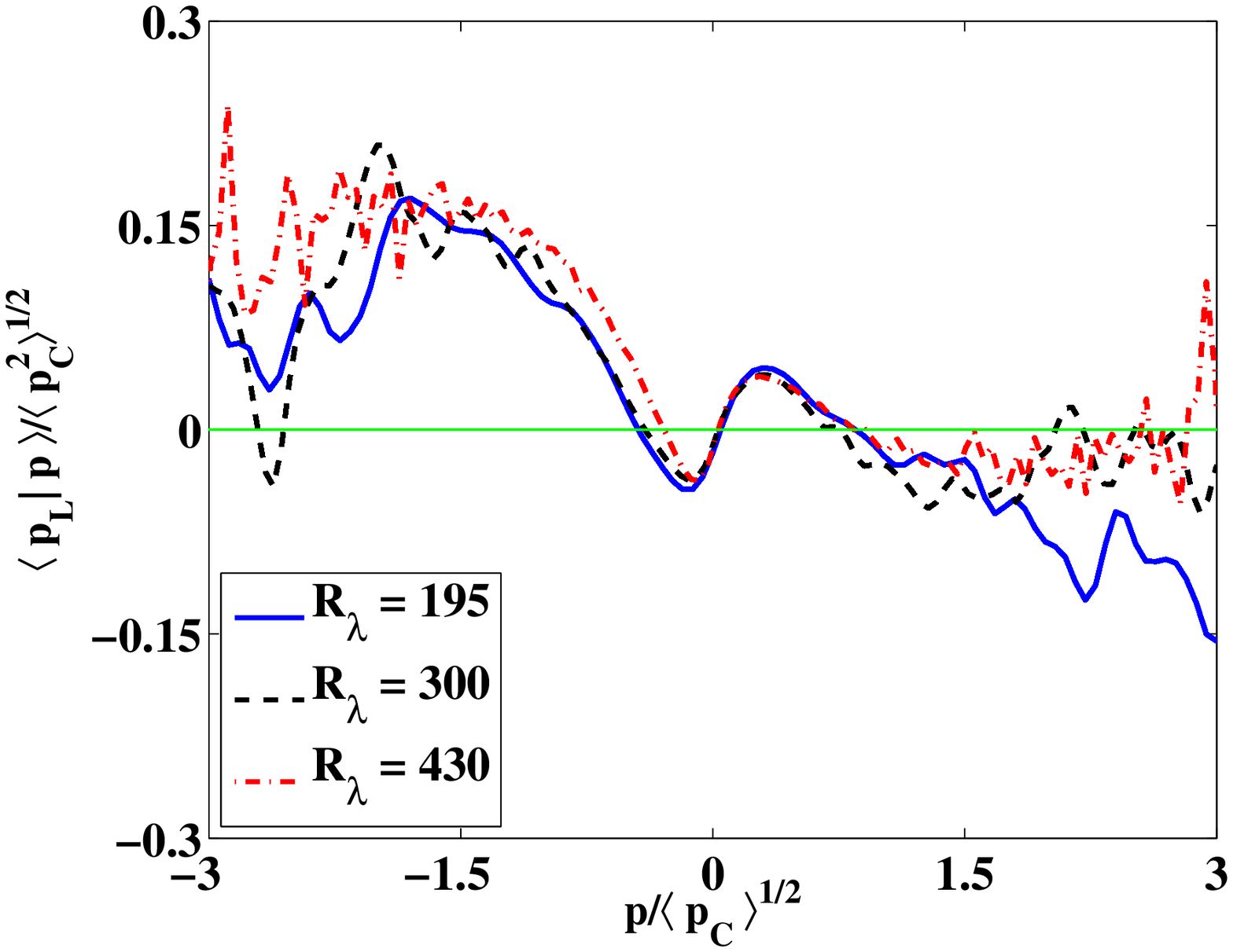}
}
\subfigure[]{
\includegraphics[width=0.48\textwidth]{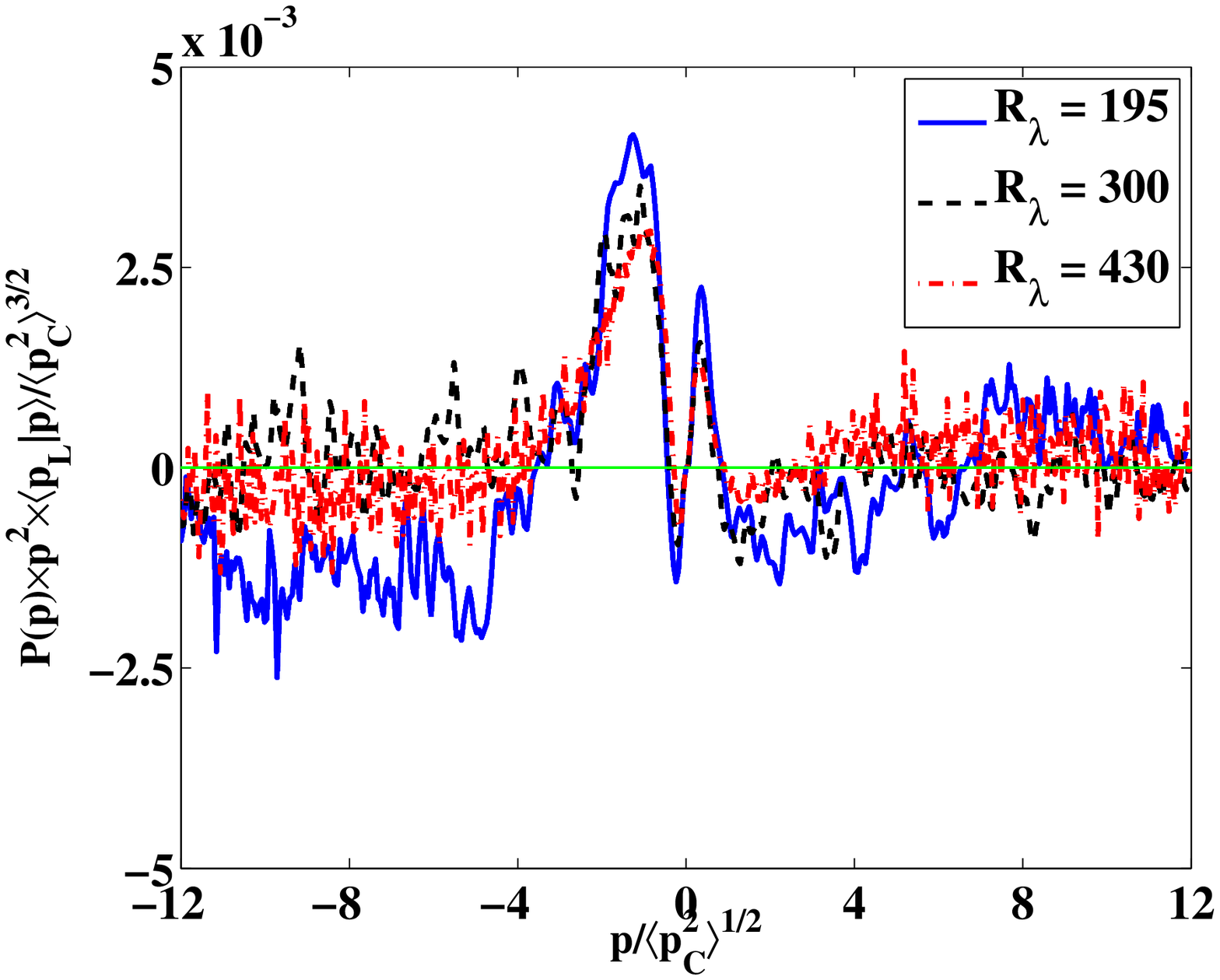}
}
\caption{The conditional PDF of $p_L$ on $p$: (a) $\langle p_L | p \rangle / \langle p_C^2 \rangle^{1/2}$ vs. $p / \langle p_C^2 \rangle^{1/2}$, 
and (b) $\mathcal{P}(p) \langle p_L | p \rangle p^2 / \langle p_C^2 \rangle^{3/2}$.
The areas under the curves in panel (b) give the normalized third moment 
$\langle p^2 p_L \rangle / \langle p_C^2 \rangle^{3/2}$, which are much smaller 
than those for $\langle p_L^2 p \rangle / \langle p_C^2 \rangle^{3/2}$ in Fig.~\ref{fig:cond_p_pL}(b).
}
\label{fig:cond_pL_p}
\end{figure*}

\subsection{Conditional averages $\langle p | p_C \rangle $ and $\langle p_C | p \rangle$}
\label{subsec:cond_avgs}

Crucial to the argument relating the third moment $\langle p^3 \rangle$ to 
the third moment of $p_C$ is the observation that the conditional averages
$\langle p | p_L \rangle$ and $\langle p_L | p \rangle$ are close to $0$.
Figure 2b of the main text shows these averages 
$( p_L, \langle p | p_L \rangle)$ and
$( \langle p_L | p \rangle, p)$, which appear as horizontal and 
vertical straight lines respectively on
the scale of the figure. Our argument is then based on identities such as:
\begin{equation}
\langle p_L^2 p \rangle = \int_{-\infty}^{\infty} \mathcal{P}(p_L) \langle p | p_L \rangle p_L^2 d p_L,
\label{eq:identity}
\end{equation}
where $\mathcal{P}(p_L)$ is the PDF of $p_L$. Equation~\eqref{eq:identity} shows
that if $\langle p_L | p \rangle = \langle p | p_L \rangle = 0$,
then, $\langle p^2 p_L \rangle = \langle p_L^2 p \rangle = 0$.

Possible deviations from zero of the moments $\langle p_L^2 p \rangle $ 
and $\langle p^2 p_L \rangle$ therefore indicate that the conditional averages
$\langle p | p_L \rangle$ and $\langle p_L | p \rangle$ are not exactly zero.
These moments can be
readily estimated from the various third moments 
$\langle p_C^m p_L^n \rangle$ with $m+n = 3$ shown in Table 2 of the main text:
\begin{equation}
\langle p^2 p_L \rangle = \langle p_C^2 p_L \rangle + 2 \langle p_C p_L^2 \rangle + \langle p_L^3 \rangle 
\end{equation}
and
\begin{equation}
\langle p_L^2 p  \rangle = \langle p_L^3 \rangle + \langle p_L^2 p_C \rangle .
\end{equation}
When compared to $\langle p_C^3 \rangle$, the value of $\langle p^2 p_L \rangle$ is approximately zero ( $\vert \langle p^2 p_L \rangle / \langle p_C^3 \rangle \vert \lesssim 2$\%), but $\langle p_L^2 p  \rangle / \langle p_C^3 \rangle \lesssim 10$\%. 
This points to a departure of
$\langle p | p_L \rangle$ from being $0$, which we explore here.

Figure~\ref{fig:cond_p_pL} shows the conditional averages of 
$\langle p | p_L \rangle$ for the three direct numerical simulation (DNS) runs discussed in this article, 
and also the integrand in Eq.~\eqref{eq:identity}.
For all three cases, the curves differ weakly but consistently from $0$. 
While for $p_L < 0$, the values of 
$\langle p | p_L \rangle$ are very small, they 
differ noticeably from $0$ on the positive $p_L$ side, with 
an approximately linear dependence on $p_L$. 
In order to examine the effect of this deviation of $\langle p|p_L \rangle$ on $\zeta = \langle p_L^2 p \rangle / \langle p_C^3 \rangle$, we non-dimensionalize the variables $p_L$ and $p$ in Fig.~\ref{fig:cond_p_pL} by $\langle p_C^2 \rangle^{1/2}$. 
In particular, in Fig.~\ref{fig:cond_p_pL}(b), we plot $\mathcal{P}(p_L) \langle p | p_L \rangle p_L^2 / \langle p_C^2 \rangle^{3/2}$. In this way, the areas under the curves in Fig.~\ref{fig:cond_p_pL}(b) give $\langle p_L^2 p \rangle / \langle p_C^2 \rangle^{3/2} = \zeta S_{p_C}$, where $S_{p_C} = \langle p_C^3 \rangle / \langle p_C^2 \rangle^{3/2}$ is the skewness of $p_C$, which depends 
weakly on the Reynolds number as shown in Table 2 of the main text. 
Fig.~\ref{fig:cond_p_pL}(b) shows that $\langle p_L^2 p \rangle / \langle p_C^2 \rangle^{3/2}$ \emph{decreases} when the Reynolds number increases. This is consistent with the observed \emph{decrease} of the value of $\zeta$ with the Reynolds number. 

As shown in Table 2 of the main text, $\zeta = 0.11$, $0.088$ and $0.066$ at $R_\lambda = 193$, $275$, and $430$, respectively. In fact, over the limited range of Reynolds number that we studied here, we observed that $\zeta / (1-\beta)$ remains well below unity: $\zeta / (1-\beta) \approx 0.64$, which ensures that the third moment of $p$, as given by Eq.~[12] in the main text, is determined by $\langle p_C^3\rangle$:
\begin{equation}
\langle p^3 \rangle = (1-\beta - \zeta) \langle p_C^3 \rangle = (1-\beta) \left( 1 - \frac{\zeta}{1-\beta} \right) \langle p_C^3 \rangle .
\end{equation}
We note that the scaling of $\langle p^3 \rangle \sim R_\lambda^2$ reported before~\cite{XPFB14}, together with the scalings $\langle p_C^3 \rangle \sim R_\lambda^3$ and $(1 - \beta) \sim R_\lambda^{-2/3}$ obtained in this work, implies that $1 - [\zeta/(1-\beta)] \sim R_\lambda^{-1/3}$. 
These predictions can only be checked by using DNS at much higher 
Reynolds numbers, and with adequate statistical resolution.

For comparison, Fig.~\ref{fig:cond_pL_p} shows the conditional average of 
$p_L$ on $p$: $\langle p_L | p \rangle$ and the integrand of $\langle p^2 p_L \rangle$.
As it was the case in Fig.~\ref{fig:cond_p_pL}, the two quantities $p$ 
and $p_L$ are normalized by
$\langle p_C^2 \rangle^{1/2}$ in the way such that the areas under the curves in Fig.~\ref{fig:cond_pL_p}(b) represent the normalized moment $\langle p^2 p_L \rangle / \langle p_C^2 \rangle^{3/2} = (\langle p^2 p_L \rangle / \langle p_C^3 \rangle) S_{p_C}$.
A systematic deviation of $\langle p_L | p \rangle$ from being $0$ is visible in Fig.~\ref{fig:cond_pL_p}(a). On the other hand, the integrand $\mathcal{P}(p) p^2 \langle p_L | p \rangle$ is noticeably non-zero only in a small range of $p$, which results in a much smaller value of $\langle p^2 p_L \rangle$ compared to $\langle p_L^2 p \rangle$.

\begin{figure*}
\centering
\subfigure[]{
\includegraphics[width=0.48\textwidth]{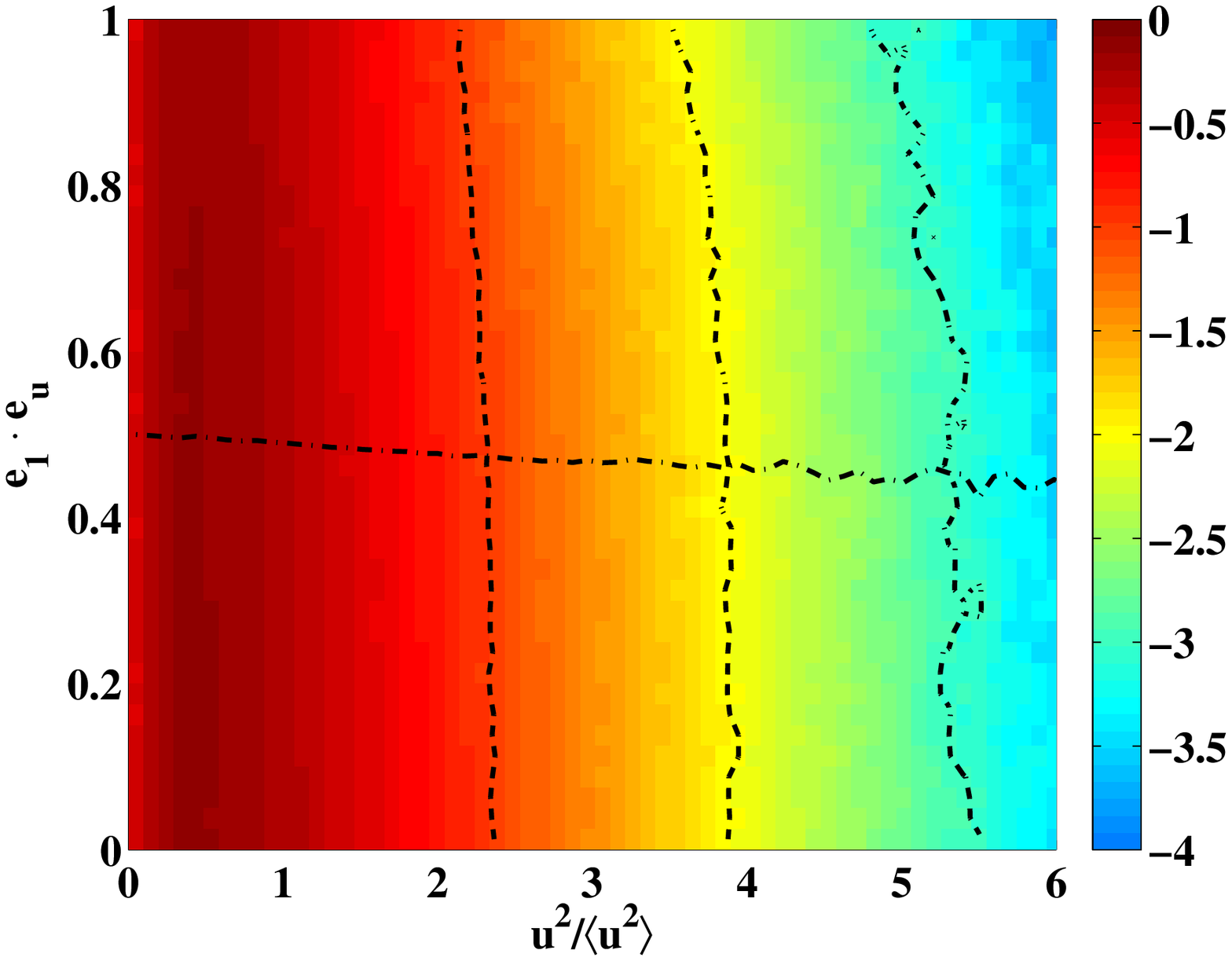}
}
\subfigure[]{
\includegraphics[width=0.48\textwidth]{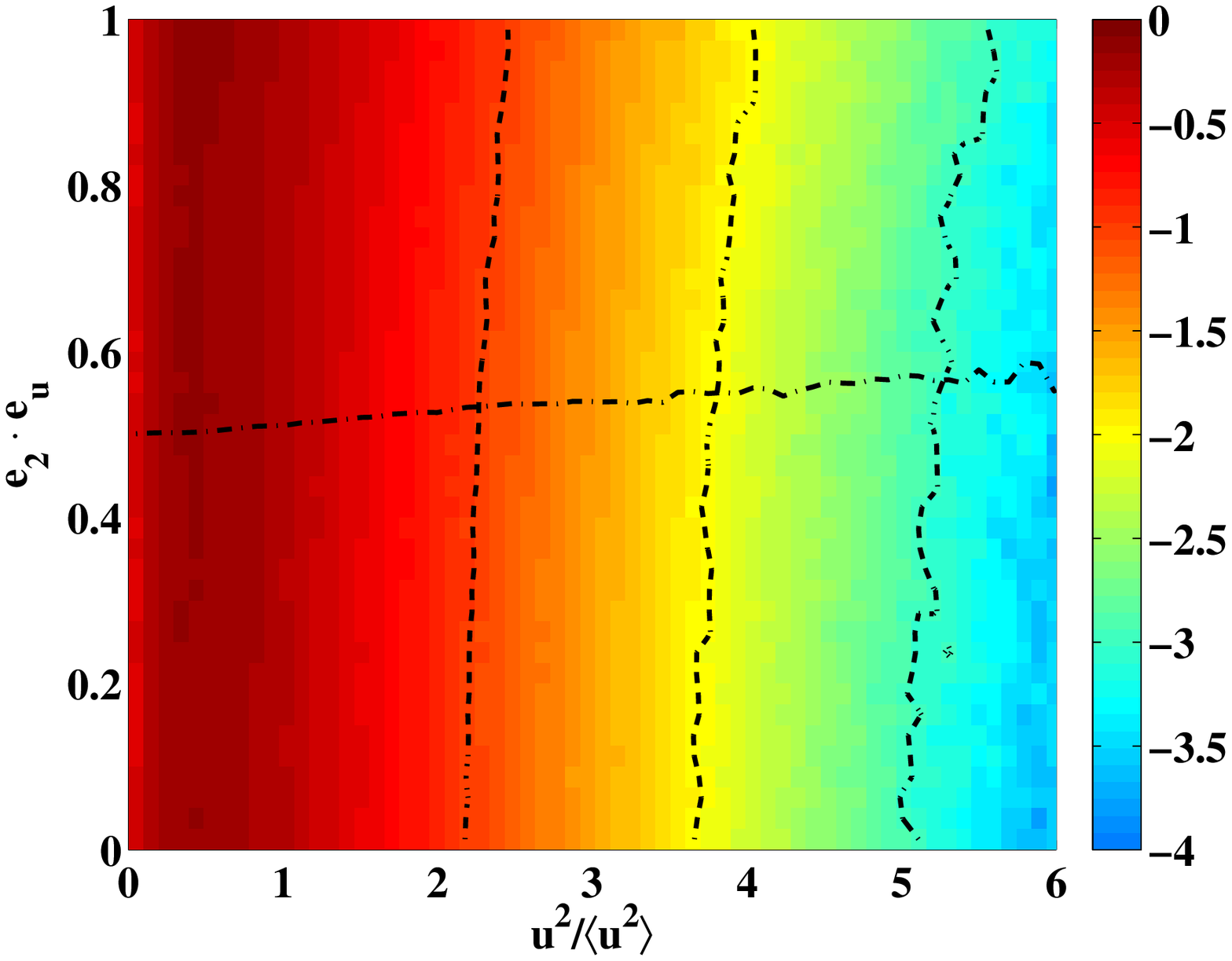}
}
\subfigure[]{
\includegraphics[width=0.48\textwidth]{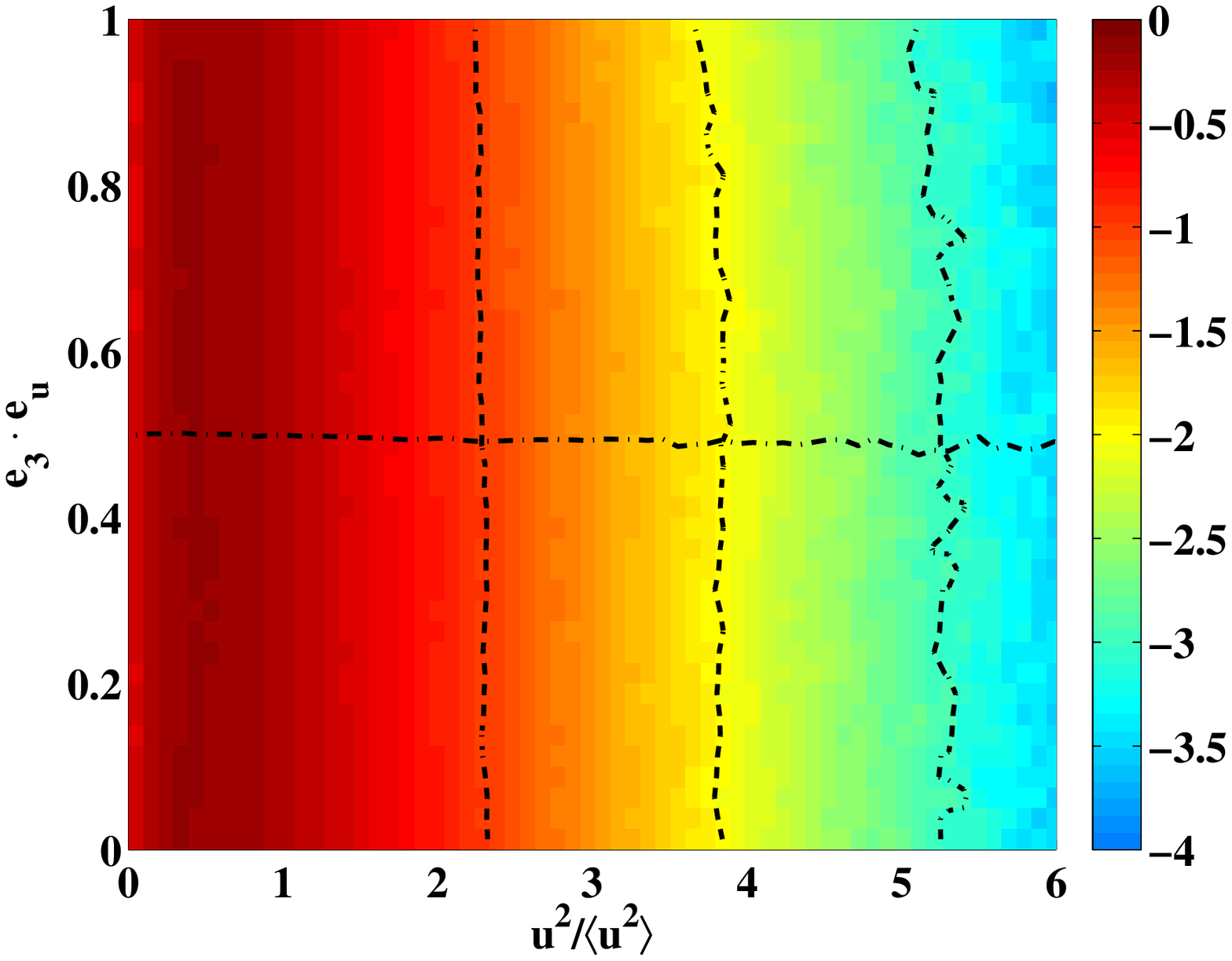}
}
\subfigure[]{
\includegraphics[width=0.48\textwidth]{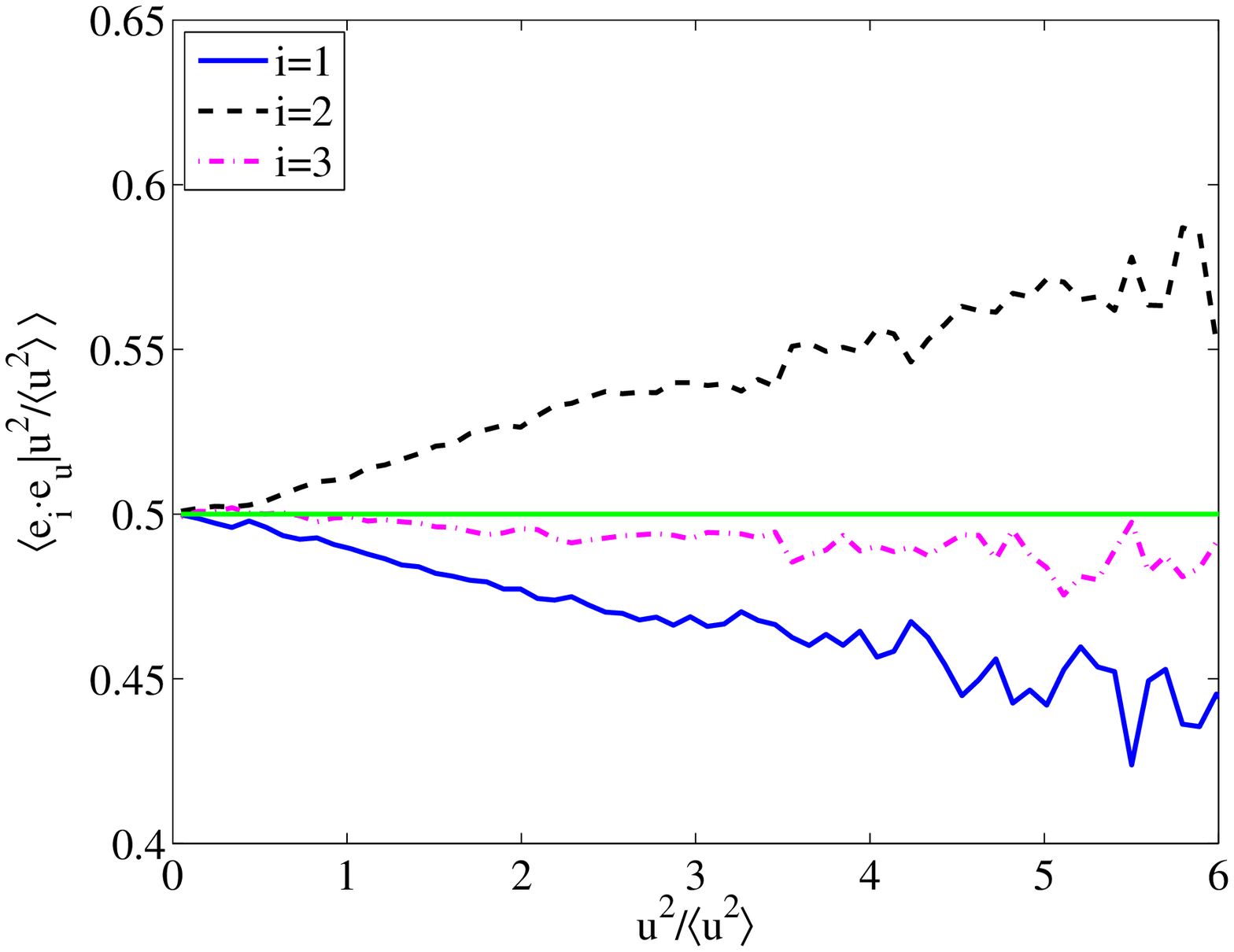}
}
\caption{Dependence of the alignment between
$\mathbf{e}_i$ and $\mathbf{e}_u$ on $\mathbf{u}^2$ at $R_\lambda = 275$.
Panels (a) (respectively (b) and (c)) show $\mathcal{P}(\mathbf{u}^2, |\mathbf{e}_u \cdot \mathbf{e}_i |)$, the joint PDF of
$\mathbf{u}^2$ and $| \mathbf{e}_u \cdot \mathbf{e}_i |$ for $i=1$
(respectively $i=2$ and $i=3$). The indicated colour coding refers to the 
decimal logarithm of the PDF, i.e., $\log_{10} \mathcal{P}(\mathbf{u}^2, |\mathbf{e}_u \cdot \mathbf{e}_i |)$. The equal-probability contours, shown as dashed
lines, are close to, but deviate from being vertical, which indicates
a systematic dependence on $\mathbf{u}^2$, especially for $i=1$ and $i=2$.
Panel (d) shows the conditional average 
$\langle |\mathbf{e}_i \cdot \mathbf{e}_u | | \mathbf{u}^2 \rangle$ vs. $\mathbf{u}^2$ (also shown
as dash-dotted lines in Panels a-c), which weakly deviates from
being constant as implied by the assumption of lack of correlation between $\mathbf{S}$ and $\mathbf{u}$.
}
\label{fig:PDF_ue_i_cond}
\end{figure*}

\begin{figure}[tb]
\centering
\includegraphics[width=0.50\textwidth]{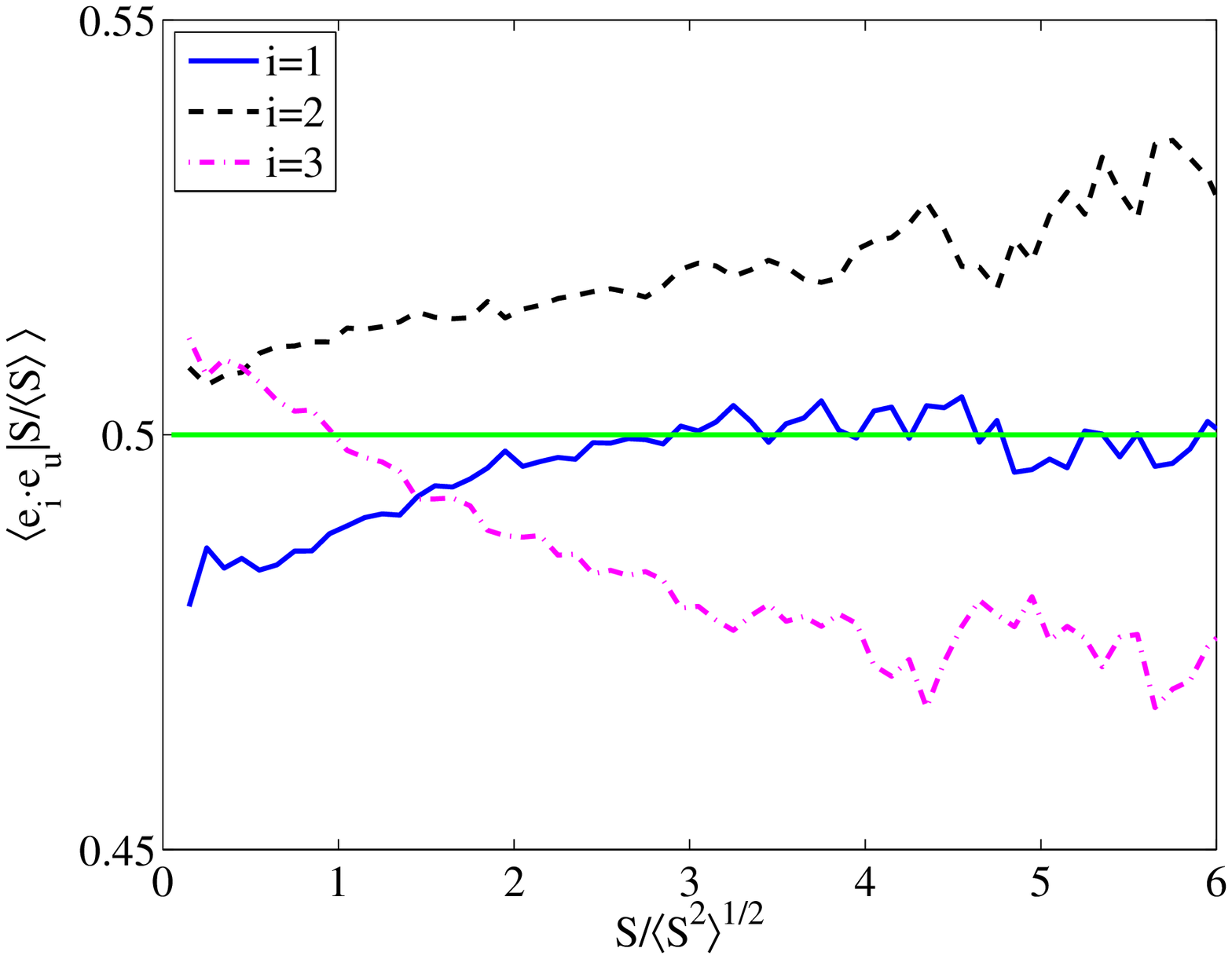}
\caption{Dependence of the alignment between
$\mathbf{e}_i$ and $\mathbf{e}_u$ on $|\mathbf{S}|$ for $ i = 1$ (full line), $i=2$ (dashed line)
and $i=3$ (dash-dotted line) at $R_\lambda = 275$. The variation of the conditional 
average of $| \mathbf{e}_i \cdot \mathbf{e}_u|$ on $ | \mathbf{S} | $
is much weaker than the dependence on $\mathbf{u}^2 $ shown in Fig.~\ref{fig:PDF_ue_i_cond}.
}
\label{fig:PDF_ue_i_cond_S}
\end{figure}

\section*{Appendix C: Lack of correlation between $\mathbf{u}$ and $\mathbf{S}$ }
\label{sec:Independence}

The results presented in the main text 
give a good indication 
that assuming $\mathbf{u}$ and $\vec{S}$ are uncorrelated
provides an appropriate first-order approximation. This is in fact corroborated by the quantitative 
agreement between the numerical results and the predictions.
Here we present further information concerning the correlations
between $\vec{u}$ and $\vec{S}$.

The first hint of a correlation between the velocity and strain was provided by 
Fig.~3 of the main text, which showed weak, but noticeable deviations from 
a uniform distribution for PDFs of the cosines of the angles between the direction of velocity, $\mathbf{e}_u$, 
and the eigenvectors of strain, especially between $\vec{e}_u$ and $\vec{e}_1$ and between $\vec{e}_u$ and $\vec{e}_2$.

It may be expected that the alignment between $\mathbf{u}$ and the
eigenvectors of $\mathbf{S}$ depends on the magnitude of $\mathbf{u}$.
Figure~\ref{fig:PDF_ue_i_cond} shows the joint probability 
distribution function between $\mathbf{u}^2$ and 
$| \mathbf{e}_u \cdot \mathbf{e}_i |$ for $i=1 $ (a), $i=2$ (b) 
and $i=3$ (c). The bending of the equal-probability contours, shown as dashed lines 
in Fig.~\ref{fig:PDF_ue_i_cond}(a-c) reveals a weak, but systematic
dependence of the alignment between $\mathbf{u}$ and $\mathbf{e}_i$ as 
a function of $\mathbf{u}^2$. Fig.~\ref{fig:PDF_ue_i_cond}(c) 
shows that the effect is considerably weaker for $i=3$ than it is for $i=1$
and $i=2$. We observe that the average of 
$| \mathbf{e}_i \cdot \mathbf{e}_u | $ conditioned on $\mathbf{u}^2$,
plotted as the dash-dotted lines in Fig.~\ref{fig:PDF_ue_i_cond}(a-c) and separately in Fig.~\ref{fig:PDF_ue_i_cond}(d),
shows clear variations as a function of
$\mathbf{u}^2$, especially for $i = 1$ and $i=2$.
The information presented in Fig.~\ref{fig:PDF_ue_i_cond} thus reveals 
that $\mathbf{u}^2$ influences not only the eigenvalues of strain, as 
shown in Figure 4 of the main text, but also the statistical properties
concerning the orientation of $\mathbf{u}$ with respect to the 
eigenvectors of $\mathbf{S}$.
This in turn induces a dependence of the third and fifth
moments of $p_C$ on $\mathbf{u}^2$ that is more complicated than
the expectation based on the lack of correlation between $\mathbf{u}$ and $\mathbf{S}$.

The results shown by Fig.~\ref{fig:PDF_ue_i_cond} 
thus show small, but systematic deviations of the statistical quantities relevant to $p_C$ from the expected dependence 
on $\vec{u}^2$.
In comparison, the dependence on the magnitude of
$\mathbf{S}$ seems to be much weaker.
Figure~\ref{fig:PDF_ue_i_cond_S} shows that the value of the average
of the cosines of the angles, $ | \mathbf{e}_u \cdot \mathbf{e}_i |$, conditioned on $| \mathbf{S} |$,
depends significantly less on $| \mathbf{S}|$: the variations shown in 
Fig.~\ref{fig:PDF_ue_i_cond_S} are of the order of $\sim 5\%$, whereas the ones shown in Fig.~\ref{fig:PDF_ue_i_cond}d are of the order of $\sim 20\%$. 
This difference points to a stronger dependence of the alignment properties on the
large scale features of the flow, than on the small scales.

The results presented in this section thus demonstrate that, 
while the results obtained in 
this work by 
assuming that $\vec{u}$ and $\vec{S}$ are uncorrelated do provide
a good approximation to the third moment of $p_C$, small, but systematic
deviations from this assumption are visible. Judging from
the present results, the dependence
on $| \vec{u} | $ seems to be generally more important than the dependence on 
$| \mathbf{S} | $.

%\bibliography{Time_irrevers}
%\bibliographystyle{pnas}

%

\end{document}